\documentclass[onecolumn]{autart}
\usepackage{}    
\usepackage[bookmarks=false]{hyperref}
\usepackage{graphicx}           
\usepackage{dcolumn}            
\usepackage{bm}                 
\usepackage{bbm}
\usepackage{graphics}           
\usepackage{epsfig}             
\usepackage{times}              
\usepackage{amsmath}
\usepackage{amssymb}
\usepackage{extarrows}
\usepackage{amsfonts}
\usepackage{mathrsfs}
\usepackage{fancyhdr}
\usepackage{color}
\usepackage{subfigure}
\usepackage{enumerate}
\usepackage{epstopdf}
\usepackage{subfiles}
  \graphicspath{{eps/}}
\usepackage [latin1]{inputenc} %
\usepackage{siunitx}
\usepackage{cite}

\usepackage{caption}
\usepackage{diagbox}

\newtheorem{theorem}{Theorem}[section] %

\newtheorem{proposition}[theorem]{Proposition}

\allowdisplaybreaks[4]
\newenvironment{pf1}{{\it Proof  of  Theorem~\ref{theorem-1}: \enspace}}{\hfill $\blacksquare$\par}
\newenvironment{pf2}{{\it Proof  of  Theorem  \ref{track-error}: \enspace}}{\hfill $\blacksquare$\par}
\newenvironment{pf3}{{\it Proof  of  Theorem  \ref{original main result}: \enspace}}{\hfill $\blacksquare$\par}
\newenvironment{pf4}{{\it Proof  of  Theorem  \ref{theorem 4}: \enspace}}{\hfill $\blacksquare$\par}

\newenvironment{pf5}{{\it Proof  of  Theorem~\ref{theorem-wellposedness}: \enspace}}{\hfill $\blacksquare$\par}
\newenvironment{pf6}{{\it Proof: \enspace}}{\hfill $\blacksquare$\par}

\begin{document}

\begin{frontmatter}

\title{Robust   synchronization   for multi-agent systems governed by PDEs  with observable and unobservable disturbances} 

\author{Yongchun Bi$^{1,2}$}\ead{yongchun.bi@my.swjtu.edu.cn},
\author{Jun Zheng$^{2}$}\ead{zhengjun2014@aliyun.com},
\author{Guchuan Zhu$^{3}$}\ead{guchuan.zhu@polymtl.ca},
\author{Jiye Zhang$^{1}$}\ead{jyzhang@swjtu.edu.cn}
\address{$^{1}$State Key Laboratory of Rail
	Transit Vehicle System, Southwest Jiaotong University,
	Chengdu 611756, Sichuan, China\\
	$^{2}$School of Mathematics, Southwest Jiaotong University,
	Chengdu 611756, Sichuan, China\\
	$^{3}${Department of Electrical Engineering, Polytechnique Montr\'{e}al, P.O. Box 6079, Station Centre-Ville, Montreal, QC, Canada H3T 1J4}
}          

\begin{abstract}
	This paper investigates  robust synchronization for multi-agent systems (MASs) governed by parabolic partial differential  equations in the presence of both observable and unobservable disturbances. Using only boundary output measurements, a disturbance observer is designed to estimate observable Dirichlet boundary disturbances while ensuring robustness of the observer error system with unobservable disturbances occurring in the domain. Using only the reference signal and local output information, distributed synchronization controllers are then constructed to enable all agents to track the reference trajectory. In particular, exponential tracking is achieved in the absence of unobservable disturbances, while robustness is preserved when additional unobservable disturbances occur during controller implementation. We further analyze the impact of unobservable Dirichlet-Robin boundary disturbances  on synchronization performance by proving the boundedness of solutions to the synchronization error system. Moreover, to characterize the influence of all disturbances, input-to-state stability (ISS) is established for the closed-loop system. For the involved systems, the generalized Lyapunov method and the recursion technique are extensively employed in the stability analysis, and the lifting technique and semigroup theory are used to prove the well-posedness. Simulation results validate the proposed control scheme, demonstrating effective disturbance estimation and rejection, robust synchronization, and the ISS properties under various scenarios.
\end{abstract}
\end{frontmatter}
\section{Introduction}
In the past decade, multi-agent systems (MASs) governed by  partial differential equations (PDEs)    have been  extensively studied in, e.g., \cite{Frihauf2011,Meurer2011,Qi2015,qi2018,Qi2019,Tang2017ijc,Qi2019ijc,zhang2024auto,wang2024auto,Ghods2012,Demetriou2013scl,Pilloni2016,Yang2017},  among which  cooperative control  has attracted particular attention due to its important engineering applications,
	such as    cooperative control of flexible spacecraft \cite{yao2024},   flexible manipulators \cite{li2023ND},  and traffic flow \cite{Zhan2025TAC}.
A key objective for addressing such problems  is to design  synchronizing
controllers  for the MASs such that the  MASs' states asymptotically
converge to a reference trajectory.
{In  the} absence of disturbances,  synchronization problem of PDEs  has been {extensively}
studied, see, {e.g.,} \cite{Yang2017,Zhan2025TAC} for state feedback  control, \cite{Qiu2023TSMCS,Yan2024TNSE} for observer-based feedback control, \cite{ZhangIJRNC} for event-triggered control, and \cite{Zhang2024TPDS} for {adaptive  neural control}. 


The inevitable presence of various disturbances in practical systems further complicates   the  synchronizing controller  design and  robustness analysis for   MASs governed by PDEs.
It is worth noting that  on one   hand,     existing results in the literature are typically developed under relatively ideal conditions where the disturbances are either observable \cite{chen2020SCL,chen2021auto,chen2021JFI,zong2021} or partially known,   e.g.,   with given bounds \cite{he2018IET,liu2025NNLS,demetriou2018,Zhao2024TNNLS}, bounded derivatives \cite{Pilloni2016,Orlov2016IFAC}, or  governed by known ordinary differential equations (ODEs) \cite{deutscher2022}.
However, in practical control scenarios, input deviations caused by actuator saturation, communication delays, or unmodeled dynamics often manifest as   unobservable disturbances without any prior information on their magnitude, dynamics,  structure, etc. and hence,   do not satisfy the   commonly  required harsh conditions, rendering the aforementioned results difficult to apply directly.  Consequently,  in the presence of such unobservable disturbances,  the  synchronization   problem of  MASs governed by PDEs   remains largely unexplored. A notable exception is \cite{Aguilar2021}, where the authors considered  unobservable in-domain and Robin boundary disturbances and  established  the input-to-state stability (ISS)  to characterize the effect of these disturbances on the overall closed-loop multi-agent system (MAS) governed by wave equations.

On the other hand, the most of the existing works primarily focus on  simple scenarios that involve only in-domain disturbances or those acting at Neumann or Robin boundaries, e.g., \cite{he2018IET,liu2025NNLS} for in-domain disturbances, \cite{chen2020SCL,chen2021auto,chen2021JFI,demetriou2018,deutscher2022,Orlov2016IFAC,Pilloni2016,zong2021} for Neumann boundary disturbances and \cite{demetriou2018,deutscher2022} for Robin boundary disturbances.
To the best of our knowledge,  for the synchronization   problem of MASs governed by PDEs, neither the active estimation and rejection of observable Dirichlet boundary disturbances nor the case involving unobservable Dirichlet boundary disturbances has yet been considered. Moreover, when  observable and unobservable disturbances (whether in-domain or boundary)  are present simultaneously, the problem becomes significantly more complex and challenging, thereby {deserving} further investigation.

For the synchronization  problem of MASs governed by PDEs in such  scenarios, two fundamental challenges exist.	
\begin{itemize}
\item
From a control design perspective, it remains {an open problem} how to achieve robust synchronization in the presence of unobservable in-domain or boundary disturbances that lack any prior information, while simultaneously reducing inter-agent communication and enabling effective estimation and rejection of observable boundary disturbances, especially Dirichlet boundary disturbances.
\item
From a stability analysis standpoint,   Dirichlet boundary disturbances introduce nonhomogeneous Dirichlet boundary terms into the tracking or synchronization error system, which are inherently difficult to handle within the framework of the classical Lyapunov stability theory without {involving the} derivative  of disturbances \cite{Mironchenko2020}.
\end{itemize}

To address  these  issues, in this paper, we investigate the  synchronization  problem of MASs governed by PDEs under a complex scenario characterized by  three types of disturbances:
\begin{enumerate}
\item[(i)]
unobservable in-domain disturbances;
\item[(ii)]  unobservable implementation disturbances at the Dirichlet and Robin boundaries; and
\item[(iii)]  observable Dirichlet boundary disturbances.
\end{enumerate}
To  ensure that all agents achieve common objectives under such  disturbances,     we propose a distributed control scheme    that relies solely on local interactions and neighbor-to-neighbor communication, and assess the stability  within the framework of a generalized Lyapunov stability theory (see \cite{Zheng2024,Zheng2025}), thereby enhancing the robustness, extensibility, and flexibility  for  MASs as indicated in \cite{mcarthur2007,Morstyn2016}. More specifically,
the present work are fourfold:
\begin{enumerate}
\item[(i)] For observable disturbances acting at the Dirichlet
boundary, a disturbance observer is designed for each agent. The proposed observer is capable of estimating these boundary disturbances even in the presence of  other unobservable   in-domain and  boundary   disturbances.
\item[(ii)]  Despite  the presence of  unobservable in-domain and Dirichlet-Robin boundary disturbances, we design   distributed synchronization controllers that rely  only on the reference signal and the output of a single neighboring agent. This design scheme not only reduces the   communication burden  among  {the agents} but also ensures the robust synchronization.
In particular,  it guarantees {the}  exponential stability   for both the tracking error and the synchronization error of the MASs with only observable  Dirichlet boundary disturbances.
\item[(iii)]  For the purpose of characterizing the influence of  disturbances on synchronization performance, we establish the  ISS  for the closed-loop system, thereby providing a framework   to quantify its  robustness.

\item[(iv)]  For stability analysis, a generalized Lyapunov method (GLM, \cite{Zheng2024,Zheng2025}), combined with the recursion technique, is extensively employed to overcome the difficulties posed by    Dirichlet boundary disturbances in the synchronization control of the
MAS. This approach enables not only the derivation of robustness estimates for the tracking and synchronization error systems but also the establishment of the  ISS  of the closed-loop system, despite the presence of various observable or unobservable disturbances. Furthermore, robustness estimates for the disturbance observer system are derived via a classical Lyapunov method when unobservable in-domain disturbances are present.

%
%
\end{enumerate}

The rest of this paper is organized as follows. Section~\ref{Sec. II} introduces the problem setting and preliminaries. In Section~\ref{Sec. III}, we present the {controller design} and the main results. The well-posedness of closed-loop system is proved in Section~\ref{IV}. The   robustness of disturbance estimation,  tracking,  and synchronization error systems, as well as the  ISS  of the closed-loop system, are analyzed in Section~\ref{V}. Numerical simulations are provided in Section~\ref{Sec. VI} to validate the theoretical findings. Finally, concluding remarks are given in Section~\ref{Sec. VII}.

\textit{Notation:}  Let  {${\mathbb{N}_0}:=\{0,1,2,\dots\}$, ${\mathbb{N}}:=\mathbb{N}_0\setminus\{0\}$, $\mathbb{R}:=(-\infty,+\infty)$, $\mathbb{R}_{\geq 0}:=[0,+\infty)$, $\mathbb{R}_{> 0}:=(0,+\infty)$, and $\mathbb{R}_{\leq 0}:=(-\infty,0]$.}

For   a domain $\Omega$ (either open or closed) in $\mathbb{R}^1$ or $\mathbb{R}^2$,   the notation   $L^2(\Omega)$ denotes the standard Hilbert space  with elements defined over $\Omega$ and the inner product $\langle \cdot,\cdot\rangle_{L^2(\Omega)}$. 
Let $H^2(\Omega) := \{v: {\Omega} \rightarrow \mathbb{R}| v$ {and all its weak partial derivatives up} to order 2 belong to  $L^2(\Omega)\}$.
Let $C\left({\Omega}\right):=\{v: {\Omega} \rightarrow \mathbb{R}|~$$v$ is continuous on $\Omega$\} and  $ \|h\|_{C(\overline{\Omega})}:=\max_{s\in  \overline{\Omega} }|h(s)|$ for  $h\in C(\overline{\Omega})$.  For $m\in \mathbb{N}$, let $C^m\left({\Omega}\right):=\{v: {\Omega} \rightarrow\mathbb{R}|~ v$ has continuous derivatives up to order $m$ on ${\Omega}$\}.  For  a normed linear space  $Y$,  let  $ C\left([0,1]; Y\right):=\{v: [0,1] \rightarrow Y|~$$v$ is continuous on $[0,1]$\} and  $ C^1\left([0,1]; Y\right):=\{v: [0,1] \rightarrow Y|~$$v$ has continuous derivatives up to order $1$ on $[0,1]$\}.   For any $\tau \in \mathbb{R}_{>0}$,  let $Q_\tau := (0,1) \times  (0,\tau)$ and $\overline{Q}_\tau$$:=$$[0,1]$$\times$$[0,\tau]$. Let $Q_\infty$$:=(0,1)\times {\mathbb{R}_{>0}}$ and $\overline{Q}_\infty:=[0,1]\times {\mathbb{R}_{\geq 0}}$.

The transposition, inverse, and determinant of a matrix $A$ are denoted by $A^{\top}$, $A^{-1}$, and $\det(A)$, respectively.
For $n \in \mathbb{N}$ and $a_i\in\mathbb{R}$, $\operatorname{diag}(a_1,\dots,a_n)$ denotes the diagonal matrix with entries  {$a_i$}   on the main diagonal and zeros elsewhere. Let   $I$ and $\mathcal{I}$  denote the  identity matrix  and the identity operator,  respectively. For a given linear operator $\mathcal{A}$, ${D}(\mathcal{A})$  denote the domain of $\mathcal{A}$.

Throughout this paper, for an integer $N\geq 2 $, the index   $i$ or $j$ is understood to range over the set   $\{1,2,\dots,N\}$ unless otherwise stated. During computation, any nonpositive index, say $m$, is interpreted as $m+N$.  For  $n \in \mathbb{N}$ and $a_i\in\mathbb{R}$, let $\prod_{i=1}^{n} a_i := a_1\times a_2 \times \cdots \times a_n$.

	\section{Problem Setting and Preliminaries}\label{Sec. II}
	
	Consider a networked system consisting of $N  $ agents, where the dynamics of the $i$-th agent are described by the following  parabolic PDE under mixed boundary conditions:
	\begin{subequations}\label{muti-agent}
		\begin{align}
			u_t^{(i)}(x, t)=&\alpha u_{x x}^{(i)}(x,t)\!-\!\!\lambda u^{(i)}(x,  t)
			+ f^{(i)}(x, t),  (x, t) \in  Q_\infty, \\
			u^{(i)}(0, t)=&q^{(i)}(t)+{U^{(i)}_0(t)},t\in\mathbb{R}_{> 0}, \label{1b} \\
			u_x^{(i)}(1, t)=&-lu^{(i)}(1, t)+{U^{(i)}_1(t)},t\in\mathbb{R}_{> 0},\label{1c} \\
			u^{(i)}(x, 0)=&u_0^{(i)}(x),x\in(0,1), \\
			y^{(i)}(t)=&u^{(i)}_x(0, t),
			z^{(i)}(t)= u^{(i)}(1, t),t\in\mathbb{R}_{> 0},\label{1e}
		\end{align}
	\end{subequations}
	where   $\alpha>0$, $\lambda>0$, and $l>0$ are   constants,  $u^{(i)}(x,t)\in \mathbb{R} $ represents the state of the $i$-th agent, $u_0^{(i)}(x)$ denotes the initial data,
	$f^{(i)}(x,t)$ represents unobservable  in-domain disturbances without any prior information on their  magnitude, dynamics, and structure,   $q^{(i)}(t)$  represents  observable Dirichlet boundary disturbances,   $y^{(i)}(t),z^{(i)}(t)$ are   measurable outputs, and  $U^{(i)}_0(t),U^{(i)}_1(t)$ are the boundary control inputs, which may introduce {additional}  unobservable boundary disturbances  {in} practical implementation  (see \eqref{control lawv} in Section~\ref{Sec. III}).

	Consider the  reference signal described by
	\begin{subequations}\label{reference_system}
		\begin{align}
			u_t^{\text{ref}}(x, t)=&\alpha u_{x x}^{\text{ref}}(x, t) -\lambda u^{\text{ref}}(x, t),  (x,t)\in Q_\infty, \\
			u^{\text{ref}}(0, t)=&r(t), t\in\mathbb{R}_{> 0}, \label{ref-2}\\
			u_x^{\text{ref}}(1, t)=&-lu^{\text{ref}}(1, t), t\in\mathbb{R}_{> 0}, \\
			u^{\text{ref}}(x, 0)=&u_0^{\text{ref}}(x),x\in(0,1),\\
			z^{\text{ref}}(t)=&u^{\text{ref}}(1,t),  x\in(0,1),
		\end{align}
	\end{subequations}
	where $r(t)$ is an arbitrarily given signal, $ z^{\text{ref}}(t)$ is the boundary output, and $u_0^{\text{ref}}(x)$ denotes the initial data.

	The main control objective is to  design  $U^{(i)}_0(t)$ and $U^{(i)}_1(t)$ by   using only  the outputs  of the reference signal and {the information of} a single neighboring agent such that
	\begin{enumerate}
		\item[(i)]
		In   the absence of   the unobservable in-domain  disturbance $f^{(i)}(x,t)$,	the  Dirichlet boundary disturbance $q^{(i)}(t)$  can always be estimated and rejected despite any other {unobservable boundary} disturbances in $U^{(i)}_0(t)$ and $U^{(i)}_1(t)$ (see \eqref{control lawv} in Section~\ref{Sec. III}), while,  in the presence of $f^{(i)}(x,t)$, the estimation error for  $q^{(i)}(t)$ remains bounded.

		\item[(ii)]     
		
		In   the absence of  unobservable in-domain and  {boundary} disturbances, all agents track the reference trajectory and maintain {an}  exponential synchronization.
		\item[(iii)]    In the presence of   unobservable in-domain and  {boundary} disturbances (see \eqref{control lawv} in Section~\ref{Sec. III}), the tracking  and    synchronization error  systems remain {bounded}, and the closed-loop system achieves the ISS.
	\end{enumerate}

	

	\section{Controller Design and Main Results}\label{Sec. III}
	In this section, we propose a distributed control design scheme for the MAS to estimate and reject observable disturbances while achieving   robust synchronization in the presence of unobservable disturbances, and then present the main results.
	\subsection{Controller Design}
	
	Following the   active disturbance rejection control scheme, which offers a framework for real-time disturbance estimation and rejection via an extended state observer \cite{Han2009TIE,guo2015TAC}, we  design an observer   to estimate the Dirichlet boundary disturbance $q^{(i)}(t)$ and then design  distributed controllers for ensuring the robust   synchronization of the MAS with various disturbances.
	
	First, we construct the following $v^{(i)}$-system to decouple $q^{(i)}(t)$  from the control action:
	%
	\begin{subequations}\label{v(x,t)}
		\begin{align}
			v_t^{(i)}(x, t)=&\alpha v_{x x}^{(i)}(x, t) - \lambda v^{(i)}(x, t), (x,t)\in Q_\infty, \\
			v^{(i)}(0, t)=&{U^{(i)}_0(t)},t\in\mathbb{R}_{> 0},\label{3b} \\
			v_x^{(i)}(1, t)=&-l v^{(i)}(1,t)+{U^{(i)}_1(t)},t\in\mathbb{R}_{> 0}, \label{3c}  \\
			v^{(i)}(x, 0)=&v_0^{(i)}(x), x\in(0,1), \\
			y_v^{(i)}(t)=&v_x^{(i)}(0, t),t\in\mathbb{R}_{> 0}, \label{3e}
		\end{align}	
	\end{subequations}
	where $y_v^{(i)}(t)$ is the measurable output and $v_0^{(i)}(x)$ denotes the initial data.
	
	Using the measured outputs $y^{(i)}(t)$ and $y_v^{(i)}(t)$, we design the following observer:
	\begin{subequations}\label{disturbances-observer}
		\begin{align}		
			\hat{q}_t^{(i)}(x, t)=&\alpha \hat{q}_{x x}^{(i)}(x, t)-\lambda \hat{q}^{(i)}(x, t), (x,t)\in Q_\infty, \\
			\hat{q}_x^{(i)}(0, t)=& y^{(i)}(t)-y_v^{(i)}(t),t\in\mathbb{R}_{> 0},\label{4b} \\
			\hat{q}_x^{(i)}(1, t)=&-l  \hat{q}^{(i)}(1, t),t\in\mathbb{R}_{> 0}, \\
			\hat{q}^{(i)}(x, 0)=&\hat{q}_0^{(i)}(x), x\in(0,1),
		\end{align}
	\end{subequations}
	where $\hat q_0^{(i)}(x)$ denotes the initial data. Then, $\hat q^{(i)}(0,t)$ can be used as an estimate of   $q^{(i)}(t)$.
	
	Now, based on  $\hat q^{(i)}(0,t)$, we design    $U^{(i)}_0(t)$ and $U^{(i)}_1(t)$   to reject the Dirichlet boundary disturbance $q^{(i)}(t)$ and achieve synchronization and reference tracking  for the MASs.
	To reduce communication burden, we employ   a distributed control scheme  where the agents are connected over a directed cycle graph. Specifically, agent $1$ receive information from agent $N$, and agent $i$ (for $i=2,\dots,N$) receive  information from its immediate predecessor, agent $i-1$.  Then, we define 
	\begin{subequations}\label{control lawv'}
		\begin{align}
			U^{(i)}_0(t)=&r(t)-\hat q^{(i)}(0,t),i=1, \dots,N,\\
			U^{(1)}_1(t)=&k_1 \left(u^{(N)}(1,t)-u^{\text{ref}}(1,t)\right),\\
			U^{(i)}_1(t)=&k_i \left(u^{(i-1)}(1,t)-u^{\text{ref}}(1,t)\right), i=2, \dots,N,
		\end{align}
	\end{subequations}
	where $ k_i \in(0,l)$ are the control gains.

	\begin{rem}
		Under the control laws given by \eqref{control lawv'},   information communication within the  MAS occurs over a   directed cycle topology.
		Notably, in practical MASs,   protocols with directed information flow incur lower communication and sensing costs compared to their undirected counterparts. Under such topologies, each agent is required to sense information from only one other agent. This minimizes the total number of communication links to $N$ for an $N$-agent system while preserving overall system connectivity~\cite{Marshall2004}.
		Owing to this   structural simplicity and high efficiency, the directed cycle topology has been successfully deployed in a wide range of practical problems within the framework of ODEs such as vehicle formations \cite{Marshall2004,Konduri2013,Rogge2008}, military applications \cite{Kumar2020}, and target circumnavigation \cite{Yu2019}. In this paper, we extend the application of directed cycle topology to the  MASs described  by PDEs.
	\end{rem}

	Note that  $U^{(i)}_0(t)$  enables the MAS to track a prescribed reference trajectory, while $U^{(i)}_1(t)$ coordinates synchronization among agents. {However,}  in practical implementations, the control inputs applied at the boundary often deviate from the ideally designed $U^{(i)}_0(t)$ and $U^{(i)}_1(t)$ due to factors such as quantization errors, signal noise, computational inaccuracies, or time delays. Taking into account this complex scenario, let $d_0^{(i)}(t)$ and $d_1^{(i)}(t)$ denote the unobservable disturbances {arising in  implementation}. Then, the actual control laws corresponding to $U^{(i)}_0(t)$ and $U^{(i)}_1(t)$ {take the form}
	\begin{subequations} \label{control lawv}
		\begin{align}
				U^{(i)}_0(t)=&r(t)-\hat q^{(i)}(0,t)+d_0^{(i)}(t),i=1, \dots,N,\\
					U^{(1)}_1(t)=&k_1 \left(u^{(N)}(1,t)-u^{\text{ref}}(1,t)\right)+{d_1^{(1)}(t)},\\
					U^{(i)}_1(t)=&k_i \left(u^{(i-1)}(1,t)-u^{\text{ref}}(1,t)\right){+d_1^{(i)}(t)},i=2, \dots,N.
				\end{align}
			\end{subequations}

				\subsection{Main Results}
				In this section, we present the main results, namely, the robustness estimates   for the estimation error $q^{(i)}(t)-\hat q^{(i)}(0,t)$,    the tracking error, and the synchronization error, as well as  the  ISS  for the closed-loop    $(u^{(i)},u^{\text{ref}},v^{(i)}, \hat q^{(i)})$-system composed of \eqref{muti-agent}--\eqref{control lawv}.
				
				To begin with, we define the   tracking  error
				\begin{align}\label{u-uref}
					\widetilde{u}^{(i)}(x,t)=u^{(i)}(x,t)-u^{\text{ref}}(x,t),
				\end{align}
				which satisfies
				\begin{subequations}\label{widetildeu}
					\begin{align}
						\!\widetilde{u}^{(i)}(x, t)=&\alpha \widetilde{u}^{(i)}_{xx}(x, t)-\lambda \widetilde{u}^{(i)}(x, t) +f^{(i)}(x, t), (x,t)\in Q_\infty, \\
						\!\widetilde{u}^{(i)}(0, t)=&q^{(i)}(t)+U^{(i)}_0(t)-r(t), t\in\mathbb{R}_{>0}, \\
						\!\widetilde{u}^{(i)}_x(1, t)=&\!-\!l\widetilde{u}^{(i)}(1, t)+U^{(i)}_1(t), t\in\mathbb{R}_{>0}, \\
						\!\widetilde{u}^{(i)}(x, 0)=&\widetilde{u}_0^{(i)}(x):=u^{(i)}_0(x)-u^{\text{ref}}_0(x),x\in(0,1).
					\end{align}
				\end{subequations}
				Substituting the control inputs given by \eqref{control lawv} into \eqref{widetildeu} yields
				\begin{subequations}\label{widetildeu3}
					\begin{align}
						\!\widetilde{u}^{(i)}(x, t)=&\alpha \widetilde{u}^{(i)}_{xx}(x, t)-\lambda \widetilde{u}^{(i)}(x, t) +\!f^{(i)}(x, t),  (x,t)\in Q_\infty, \\
						\!\widetilde{u}^{(i)}(0, t)=&q^{(i)}(t)+d^{(i)}_0(t)-\hat q^{(i)}(0,t), t\in\mathbb{R}_{>0},\label{eq-12b} \\
						\!\widetilde{u}^{(i)}_x(1, t)=&\!-\!l\widetilde{u}^{(i)}(1, t)\!+k_i\widetilde u^{(i-1)}{(1,t)} +d^{(i)}_1(t),   t\in\mathbb{R}_{>0}, \label{12c}\\
						\!\widetilde{u}^{(i)}(x, 0)=&\widetilde{u}_0^{(i)}(x),x\in(0,1).
					\end{align}
				\end{subequations}

				The  synchronization  error between   two agents is defined by
				\begin{align*}
					\widetilde{u}^{(i,j)}(x,t):=u^{(i)}(x,t)-{u^{(j)}(x,t)}
						=	\widetilde{u}^{(i)}(x,t)-	\widetilde{u}^{(j)}(x,t),   i,j=1,\dots,N.
					\end{align*}
					For notational simplicity, for any function $h$ with   index $i$, let
					\begin{align*}
						h^{(i,j)}:= h^{(i)}-h^{(j)}.
					\end{align*}
					Then, it holds that
					\begin{subequations}\label{widetildeu-i-j}
						\begin{align}
							\widetilde{u}^{(i,j)}(x, t)=&\ \alpha \widetilde{u}^{(i,j)}_{xx}(x, t)-\lambda \widetilde{u}^{(i,j)}(x, t) +f^{(i,j)}(x, t),(x,t)\in Q_\infty, \\
							\widetilde{u}^{(i,j)}(0, t)=&q^{(i,\!j)}(t)+\!d^{(i,j)}_0(t) - {\hat q^{(i,j)}(0,t)}, t\in\mathbb{R}_{>0},  \\
							\widetilde{u}^{(i,j)}_x(1, t)=& - l\widetilde{u}^{(i,j)}(1, t)+d^{(i,j)}_1(t)  + k_i\widetilde u^{(i-1)}(1,t) -k_j\widetilde u^{(j-1)}(1,t), t \in \mathbb{R}_{>0},   \\
							\widetilde{u}^{(i,j)}(x, 0)=&{\widetilde{u}_0^{(i,j)}(x)},x\in(0,1).
						\end{align}
					\end{subequations}
						
						Throughout this paper, we assume that
						$f^{(i)}\in C(\overline{Q}_\infty)$, $d_0^{(i)},d^{(i)}_1,q^{(i)}, r \in C^1( \mathbb{R}_{\geq 0})$, and  $ \widetilde u^{(i)}_0, u^{\text{ref}}_0,   v^{(i)}_0\in H^2(0,1)\cap C([0,1])$
						satisfying  the following compatibility conditions:
						\begin{subequations}\label{CC}
							\begin{align}
								\widetilde{u}^{(i)}_0(0)= &q^{(i)}(0)+d^{(i)}_0(0)-\hat q^{(i)}_0(0),  \\
								\widetilde{u}^{(i)}_{0x}(1)=& - l\widetilde{u}^{(i)}_0(1)+k_i\widetilde u^{(i-1)}_0{(1)} +d^{(i)}_1(0), \\
								u^{\text{ref}}_0(0)=&r(0),
								u_{0x}^{\text{ref}}(1)=-lu_0^{\text{ref}}(1), \\
								v^{(i)}_0(0)=&{U^{(i)}_0(0)},
								v_{0x}^{(i)}(1)=-l v_0^{(i)}(1)+{U^{(i)}_1(0)}.
							\end{align}
						\end{subequations}	
						
						The first main result is concerned with the
						{boundedness}  of the {estimation} error  {$q^{(i)}(t)-\hat q^{(i)}(0,t)$}.
						\begin{theorem} \label{theorem-1}
							The {estimation} error   $q^{(i)}(t)-\hat q^{(i)}(0,t)$ admits the following estimate with any $\sigma\in \left(0,\lambda\right)$  for all $t \in \mathbb{R}_{>0}$:
							\begin{align}\label{q-hatw}
								\left|q^{(i)}(t)- \hat q^{(i)}(0,t)\right|\leq \left\|q^{(i)}(t)-\hat q^{(i)}(\cdot,t)\right\|_{C([0,1])}
								\leq    e^{ - \sigma   t}  \left\|u_0^{(i)}-v_0^{(i)}-\hat q_0^{(i)}\right\|_{C([0,1])} + \frac{1}{\lambda-\sigma}\left\|f^{(i)}\right\|_{C(\overline{Q}_t)}.
							\end{align}
							
						\end{theorem}
						\begin{rem}
							The estimate \eqref{q-hatw}   confirms  that the estimation error   converges exponentially to zero    in the absence of the in-domain disturbances $f^{(i)}$, while   remaining bounded in their presence.
						\end{rem}

						Building on the boundedness of the estimation error, the second main result establishes the robustness of the tracking error $\widetilde{u}^{(i)}$-system~\eqref{widetildeu3}  under  the control laws given by \eqref{control lawv}.
						\begin{theorem}\label{track-error}
							Under the control laws given by \eqref{control lawv},
							the  {tracking  error $\widetilde{u}^{(i)}$-system~\eqref{widetildeu3}}  admits  the  estimate    in the max-norm   for all $ t\in \mathbb{R}_{> 0}$:
							\begin{align}\label{tildeu-es}
								\left\|\widetilde{u}^{(i)}(\cdot,t)\right\|_{C([0,1])}
								\leq\widetilde{\mathcal{C}}_i\mathcal{D}(t) +\widetilde{\mathcal{C}}_ie^{-\sigma t}\left(\max_{i} \left\|\widetilde{u}^{(i)}_0\right\|_{C([0,1])}+  \max_{i}\left\|u^{(i)}_0\!-v^{(i)}_0 -\hat q^{(i)}_0\right\|_{C([0,1])}\right),
							\end{align}
							where
							\begin{subequations}
								\begin{align*}
									\mathcal{D}(t):= &  \frac{2}{\lambda-\sigma}\max_{i}\left\|{f}^{(i)}\right\|_{C(\overline{Q}_t)}
									+\max_{i}\left\|{d^{(i)}_0}\right\|_{C([0,t])}+\frac{1}{l}\max_{i} \left\|{d^{(i)}_1}\right\|_{C([0,t])},\notag\\
									\mathcal{C}_i:=&  \frac{ \mathcal{M}_{i}}{  1-\mathcal{N} },\notag\\
									 \widetilde{\mathcal{C}}_i:=&   1+\frac{k_i}{l}{\mathcal{C}}_{i-1},\\ \mathcal{N}:=&   \frac{\prod_{i=1}^{N}k_{i}}{{|l|^{N}}}, \notag\\
									\mathcal{M}_i:= & 1+\sum_{m=1}^{i}\frac{\prod_{r=0}^{m-1}k_{i-r}}{l^{m}}+\sum_{m=i+1}^{N-1}\frac{\left(\prod_{r=0}^{i-1}k_{i-r}\right)\left(\prod_{s=0}^{m-i-1}k_{N-s}\right)}{l^{m}},
								\end{align*}
								and $\sigma\in \left(0,\lambda\right)$ is an arbitrary  constant.
							\end{subequations}
						\end{theorem}
						
						The third main  result shows the  robustness of  the synchronization error $\widetilde{u}^{(i,j)}$-system~\eqref{widetildeu-i-j} under  the control laws given by \eqref{control lawv}.		
						\begin{theorem}\label{original main result}
							Under the control laws given by \eqref{control lawv},
							the synchronization error $\widetilde{u}^{(i,j)}$-system~\eqref{widetildeu-i-j}   admits  the   estimate  in the max-norm  for all $t \in \mathbb{R}_{>0}$:
							\begin{align}\label{tildeu-i-j-es}
								\left\|\widetilde{u}^{(i,j)}(\cdot,t)\right\|_{C([0,1])}
								\leq& \widetilde{\mathcal{C}}_j \widetilde{\mathcal{D}}(t)+\widetilde{\mathcal{C}}_je^{-\sigma t}\bigg(\max_{i,j}\left\|\widetilde{u}^{(i,j)}_0\right\|_{C([0,1])}+ \max_{i,j}\left\|u^{(i,j)}_0\!-v^{(i,j)}_0 -\hat q^{(i,j)}_0\right\|_{C([0,1])}\bigg)\notag\\
								& +\widetilde{\mathcal{H}}_{i,j}\widetilde{\mathcal{C}}_i\left\{\mathcal{D}(t)+  e^{-\sigma t}\left( \max_{i} \left\|\widetilde{u}^{(i)}_0\right\|_{C([0,1])}+ \max_{i}\left\|u^{(i)}_0\!-v^{(i)}_0 -\hat q^{(i)}_0\right\|_{C([0,1])}\right)\right\},
							\end{align}
							where
							\begin{subequations}
								\begin{align*}
									\widetilde{\mathcal{D}}(t):=&   \frac{2}{\lambda-\sigma}\max_{i,j}\left\|{f}^{(i,j)}\right\|_{C(\overline{Q}_t)}
									+\max_{i,j}\left\|{d^{(i,j)}_0}\right\|_{C([0,t])}
									+\frac{1}{l}\max_{i,j} \left\|{d^{(i,j)}_1}\right\|_{C([0,t])},  \\
									\mathcal{H}_{i,j}:=&\frac{|k_i-k_j|}{l}
									+
									\sum_{m=1}^{j}
									\frac{(\prod_{r=0}^{m-1}k_{\,j-r})\,|k_{\,i-m}-k_{\,j-m}|}{ l^{\,m+1}}+
									\sum_{m=j+1}^{N-1}
									\frac{\mathcal{K}_{j,m}
										|k_{\,i-m+N}-k_{\,j-m+N}|}{ l^{\,m+1}},\\
									\mathcal{K}_{j,m}:=& \left(\prod_{r=0}^{j-1}k_{\,j-r} \right) \left(\prod_{s=1}^{\,m-j}k_{\,N+1-s} \right),\\
									\widetilde{\mathcal{H}}_{i,j}:=&\frac{k_j\mathcal{H}_{i-1,j-1}}{l(1-\mathcal{N})}+\frac{|k_i-k_j|}{l},
								\end{align*}
							\end{subequations}
							and $\sigma, \mathcal{N}$, $\widetilde{\mathcal{C}}_i$,  and $\mathcal{D}(t)$   are the same as in   Theorem~\ref{track-error}.

						\end{theorem}
						\begin{rem}
							The estimates   \eqref{tildeu-es} and  \eqref{tildeu-i-j-es} show that, in the absence {of unobservable} disturbances, i.e., $f^{(i)}\equiv d_0^{(i)}\equiv d_1^{(i)}\equiv0 $, all agents  track the reference signal and achieve exponential synchronization.   When the unobservable disturbances are bounded,  {the  $\widetilde{u}^{(i)} $-system~\eqref{widetildeu3}} and {the $\widetilde{u}^{(i,j)}$-system~\eqref{widetildeu-i-j}}   remain {bounded. Moreover,} as the  bounds of unobservable disturbances decrease, the   bounds of $\widetilde u^{(i)}$ and  $\widetilde{u}^{(i,j)} $   decrease.
						\end{rem}
						
						The fourth result  is concerned with  the ISS with respect to (w.r.t.) $f^{(i)},d_0^{(i)},d_1^{(i)}, r$, and $q^{(i)}$ for the closed-loop system consisting of \eqref{muti-agent}--\eqref{disturbances-observer} under the control laws  given by~\eqref{control lawv}.
						\begin{theorem}\label{theorem 4}
							Under the control laws given by \eqref{control lawv},
							the closed-loop $(u^{(i)},u^{\text{ref}},v^{(i)}, \hat q^{(i)})$-system composed of \eqref{muti-agent}--\eqref{disturbances-observer} is input-to-state stable (ISS)  in the max-norm,  having  the following  estimate for all $t \in \mathbb{R}_{>0}$:
							\begin{align}\label{closedloop}
								& \left\| u^{(i)}(\cdot, t) \right\|_{C([0,1])} + \left\|  u^{\text{ref}}(\cdot, t) \right\|_{C([0,1])} + \left\|  v^{(i)}(\cdot, t) \right\|_{C([0,1])} + \left\| \hat{q}^{(i)}(\cdot, t) \right\|_{C([0,1])} \notag\\
								\le  & \left( 1  +2 \widetilde{\mathcal{C}}_i \right) e^{-\sigma t}\left(  2 \max_{i} \left\| u_0^{(i)} \right\|_{C([0,1])} + \left\| u_0^{\text{ref}} \right\|_{C([0,1])}\right)  +  \left( 4   + 2\widetilde{\mathcal{C}}_i \right)   e^{-\sigma t}\left(\! \max_{i} \left\| v_0^{(i)} \right\|_{C([0,1])} \!+ \max_{i} \left\| \hat q_0^{(i)} \right\|_{C([0,1])} \right) \notag\\
								& + \!3\left\|   r \right\|_{C([0,t])} \!+ 2\left\|   q^{(i)}\right \|_{C([0, t])}\!+ \!\left( 1 \!+2 \widetilde{\mathcal{C}}_i \right) \!\mathcal{D}(t) ,  
							\end{align}
							where 	$\widetilde{\mathcal{C}}_i$,  $\mathcal{D}(t)$, and $\sigma$ are the same as in    Theorem~\ref{track-error}.
						\end{theorem}
						\begin{rem}
							The estimate \eqref{closedloop} shows that the MAS~\eqref{muti-agent} achieves robust synchronization and robust tracking of the reference system~\eqref{reference_system}, where the synchronization error decays exponentially   and is ultimately bounded by the magnitudes of the reference input  $r(t)$, the observable disturbances   $q^{(i)}(t)$, and the unobservable   disturbances described by $\mathcal{D}(t)$.
						\end{rem}
						\begin{rem}
							Unlike the existing results on synchronization of MASs governed by PDEs considered in Hilbert spaces~\cite{Aguilar2021,chen2020SCL,chen2021auto,chen2021JFI,demetriou2018,deutscher2022,he2018IET,liu2025NNLS,Orlov2016IFAC,Pilloni2016,zong2021,Zhao2024TNNLS},  the estimates   \eqref{q-hatw}, \eqref{tildeu-es}, \eqref{tildeu-i-j-es}, and \eqref{closedloop} are given  in  continuous functional spaces. This leads to  robustness  estimates in the max-norm, namely, a point-wise estimate  for each system, thereby   providing  stronger   characterizations of the estimation, tracking, and synchronization performance for the MASs.
						\end{rem}
						\begin{rem}{Note that} the compatibility conditions given by \eqref{CC} are only required for the well-posedness of classical solutions (see Section~{IV}) and can be removed if a weak or mild solution is considered. In that case, a standard density argument (see \cite[{p. 25,  Lemma 1.4.2}]{Mironchenko:2023b}) can be used to establish the robustness estimates for each system.
						\end{rem}		
						
						%
						\section{Well-Posedness Analysis}\label{IV}
						In this section, we establish the well-posedness of all systems under consideration, including the $\widetilde u^{(i)}$-system~\eqref{widetildeu3} and the closed-loop system composed of \eqref{muti-agent}--\eqref{disturbances-observer} under the control laws given by \eqref{control lawv}.
						\subsection{Well-Posedness Analysis for the $\widetilde{u}^{(i)}$-System~\eqref{widetildeu3}} 	
						In this subsection, we employ the lifting technique and semigroup theory of linear operators to establish the well-posedness of the $\widetilde{u}^{(i)}$-system~\eqref{widetildeu3}, which is stated as the following theorem.
						\begin{theorem}\label{theorem-wellposedness}
							The	$\widetilde{u}^{(i)}$-system~\eqref{widetildeu3}  admits a unique solution 	$\widetilde{u}^{(i)} \in C(\mathbb{R}_{\geq 0};H^2(0,1))\cap C^1(\mathbb{R}_{\geq 0};L^2(0,1))$.
						\end{theorem}
						
						Before proving this theorem, we rewrite the equations of all $\widetilde u^{(i)}$-systems \eqref{widetildeu3} into a unified form. Specifically, let
						\begin{align*}
							\widetilde\nu:=&\begin{pmatrix}\widetilde u^{(1)}, \dots, \widetilde u^{(N)}\end{pmatrix}^{\top},\notag\\
							 \widetilde\nu_0 :=&\begin{pmatrix}\widetilde u_0^{(1)} , \dots, \widetilde u_0^{(N)} \end{pmatrix}^{\top},\notag\\
							{F} :=&\begin{pmatrix}f^{(1)}, \dots, f^{(N)}\end{pmatrix}^{\top}, \notag\\
							 {D}:= &\begin{pmatrix}  d^{(1)}_1, \dots,  d^{(N)}_1   \end{pmatrix}^{\top},\\
							Q:=& \begin{pmatrix}\!q^{(1)}+d^{(1)}_0\!-\hat q^{(1)}(0,t), \dots, q^{(N)}+d^{(N)}_0 \!-\hat q^{(N)}(0,t)\!\end{pmatrix}^{\!\top}\!\!\!,
						\end{align*}
						and define the matrix 
						\begin{align*}
							B   := [b_{ij}]_{N \times N}:=
							\begin{pmatrix}
								-l & 0 & 0 & \cdots & 0 & k_1 \\
								k_2 & -l & 0 & \cdots & 0 & 0 \\
								0 & k_3 & -l & \cdots & 0 & 0 \\
								\vdots & \vdots & \ddots & \ddots & \vdots & \vdots \\
								0 & 0 & \cdots & k_{N-1} & -l & 0 \\
								0 & 0 & \cdots & 0 & k_N & -l
							\end{pmatrix}.
						\end{align*}
						Then,    we get
						\begin{subequations}\label{mathcalu}
							\begin{align}
								\!\!\!\!\!	\widetilde\nu_t(x,t)=&\alpha\widetilde\nu_{xx}(x,t)-\lambda\widetilde\nu(x,t)+F(x,t),  (x,t)\in Q_\infty,\\
								\!\! \!\!\!\widetilde\nu(0,t)=&Q(t),t\in \mathbb{R}_{>0},\\
								\!\!\!\!\!\widetilde\nu_x(1,t)=&B\widetilde\nu(1,t)+D(t),t\in \mathbb{R}_{>0},\\
								\!\!\!\!\!\widetilde\nu(x,0)	=&\widetilde\nu_0(x),x\in(0,1).
							\end{align}
						\end{subequations}
						
						Next,	we
						employ a lifting technique to convert  \eqref{mathcalu} with nonhomogeneous boundary conditions into a system with homogeneous ones and rewrite it in an abstract form. Indeed, letting
						 	\begin{align*}
							 H:=  \begin{pmatrix}  h^{(1)}, \dots,  h^{(N)}\end{pmatrix}^{\top} :=Q+(x^2-x)\left(B Q+ D\right)
							 \end{align*}
							 and
							\begin{align*}
							\Phi:= \begin{pmatrix}  \phi^{(1)}, \dots,  \phi^{(N)}\end{pmatrix}^{\top}:=\widetilde\nu-H,
							 \end{align*}
						system~\eqref{mathcalu}  can be written as
						\begin{subequations}\label{phi}
							\begin{align}
								\Phi_t(x,t)=&\alpha\Phi_{xx}(x,t)+F(t,\Phi),(x,t)\in Q_\infty,\\
								\Phi(0,t)=&0, t\in \mathbb{R}_{>0},\label{14a}	\\
								\Phi_x(1,t)=&B\Phi(1,t), t\in \mathbb{R}_{>0},\label{14b}\\
								\Phi(x,0)=&\Phi_0(x):=\widetilde\nu_0(x)-H(x,0),x\in(0,1).
							\end{align}
						\end{subequations}
						
						Let $\Psi :=\begin{pmatrix}\psi^{(1)}, \dots, \psi^{(N)}\end{pmatrix}^{\top}$.	Define the  operator $\mathcal{A}:{D}(\mathcal{A})\subset X\rightarrow X:=(L^2(0,1))^N$   by
						\begin{align*}
							\mathcal{A}{\Psi} := 	\operatorname{diag}\left(   {\alpha  }(\psi^{(1)})^{\prime\prime}, \dots, {\alpha  } (\psi^{(N)})^{\prime\prime}   \right)
						\end{align*}
						with
						\begin{align*}
							{D}(\mathcal{A}): = \Big\{\Psi \in (H^2(0,1))^N~\big|~ \Psi (0) = 0,  \Psi^\prime (1) = B\Psi (1)\Big\}.
						\end{align*}
						Then,   system~\eqref{phi}, and thus system~\eqref{mathcalu}, can be  written as 	
							\begin{align}\label{Phi}
								\dot{\Phi} =   {\mathcal{A}}\Phi+ \mathcal{F}(t,\Phi),	\ \
								\Phi(0)=  \Phi_0,
							\end{align}
						where 	 		$\mathcal{F}(t,\Phi):=-\lambda\Phi+ F - H_t+\alpha H_{xx}-\lambda H$. Hence, it suffices to prove the existence of {a solution} to  system~\eqref{Phi}.
						\begin{proposition}\label{matchala}
							System~\eqref{Phi} admits a  unique  (classical) solution  $ \Phi\in C(\mathbb{R}_{\geq 0};{D(\mathcal{A}))}\cap C^1(\mathbb{R}_{>0};X)$.
						\end{proposition}
						\begin{pf6}
							By virtue of the Lumer-Phillips theorem \cite[p. 14, Theorem 4.3]{pazy1983}, we proceed {with the} proof in  two steps.
							
							\textbf{Step 1:} Show that the operator $\mathcal{A}$ is dissipative. Indeed, by the definition  of $\mathcal{A}$ and the   inner product, together with integration by parts, \eqref{14a}, and \eqref{14b}, we obtain
							\begin{align}\label{aphi}
								\langle \mathcal{A}  \Phi,  \Phi\rangle_{X}
								= & \sum_{i=1}^N \int_0^1 \alpha (\phi^{(i)})^{\prime\prime}  \phi^{(i)} \text{d} x \notag\\
								= & \sum_{i=1}^N \alpha \phi^{(i)}(1, t)(\phi^{(i)}(1, t))^{\prime}-\sum_{i=1}^N \alpha \phi^{(i)}(0, t)(\phi^{(i)}(0, t))^{\prime}  -\sum_{i=1}^N \alpha \int_0^1\left((\phi^{(i)})^{\prime}\right)^2 \text{d} x \notag\\
								= & \sum_{j=1}^N \sum_{i=1}^N \alpha b_{i j} (\phi^{(j)}(1, t))^2-\sum_{i=1}^N \alpha \int_0^1\left((\phi^{(i)})^{\prime}\right)^2 \text{d} x .
							\end{align}
							{Substituting a variation of} Agmon's  inequality (see \cite[Lemma 1]{Zheng2018}) {of the form}
							\begin{align*}
								(\phi^{(i)}(1, t))^2\leq 2 \int_0^1 (\phi^{(i)})^2 \text{d} x+\int_0^1\left((\phi^{(i)})^{\prime}\right)^2 \text{d} x
							\end{align*}
							into \eqref{aphi} yields
							\begin{align*}
								\langle \mathcal{A}  \Phi,  \Phi\rangle_{X}
								\leq & \sum_{j=1}^N \sum_{i=1}^N \alpha b_{i j}\left(2 \int_0^1 (\phi^{(j)})^2 \text{d} x+\int_0^1\left((\phi^{(j)})^{\prime}\right)^2 \text{d} x\right)  -\sum_{i=1}^N \alpha \int_0^1\left((\phi^{(i)})^{\prime}\right)^2 \text{d} x \\
								= & 2 \alpha \sum_{j=1}^N\sum_{i=1}^N b_{i j} \int_0^1 (\phi^{(j)})^2 \text{d} x  +\alpha\sum_{j=1}^N\left( \sum_{i=1}^N b_{i j}-1\right) \int_0^1\left((\phi^{(j)})^{\prime}\right)^2 \text{d} x\\
								=&2 \alpha\sum_{i=1}^{N-1}\left(-l+k_{i+1}\right) \int_0^1 (\phi^{(i)})^2 \text{d} x+2 \alpha\left(-l+k_{1}\right) \int_0^1 (\phi^{(N)})^2 \text{d} x\notag\\
								&+\alpha\sum_{i=1}^{N-1}\left(-l+k_{i+1}-1\right)\int_0^1\left((\phi^{(i)})^{\prime}\right)^2 \text{d} x+\alpha\left(-l+k_1-1\right)\int_0^1\left((\phi^{(N)})^{\prime}\right)^2 \text{d} x.
							\end{align*}
							Since $k_i \in (0,l)$ for all $i$, we have $-l + k_{i+1} < 0$ and $-l + k_1 < 0$. Combining this with $\alpha > 0$, we obtain
							$\langle \mathcal{A}\Phi, \Phi \rangle_X \leq 0$.
							Thus, $\mathcal{A}$ is dissipative.

							\textbf{Step 2:} Prove that the operator $  {\mathcal{I}} -   {\mathcal{A}}$ is surjective. That is, for any  $\Gamma \in X$, there exists $\Phi\in {D}(\mathcal{A})$ such that
							$\left(\mathcal{I}-  \mathcal{A}  \right) \Phi=  \Gamma $.
							Indeed,  from the definition of  $ \mathcal{A}$, it follows that
							\begin{subequations}\label{equation-A}
								\begin{align}
									-\alpha  	 {\Phi}^{\prime\prime} +{\Phi}=&\Gamma,\\
									{\Phi}(0)=&0, \\
									{\Phi}^{\prime}(1)=&B	 {\Phi}(1). \label{20c}
								\end{align}
							\end{subequations}
							We now turn to consider the solution to the  ODE~\eqref{equation-A}. We first construct a particular solution that satisfies the left boundary condition. Specifically, we look for a solution of the form
							\begin{align}\label{phipde}
								{\Phi}_p(x):=-\int_0^x \frac{\sinh (\gamma(x-s) )}{\alpha \gamma} \Gamma(s) \text{d}s,
							\end{align}
							where $\gamma:=\sqrt{\frac{1}{\alpha}}$. It can be directly verified that ${\Phi}_p$ satisfies
							\begin{align*}
								-\alpha {\Phi}^{\prime\prime}_p +{\Phi}_p =\Gamma,\ \
								{\Phi}_p(0) = 0.
							\end{align*}
							Therefore, the general solution to \eqref{equation-A} is given by
							\begin{align}\label{phi-ex}
								{\Phi}(x)={\Phi}_p(x)+{\Phi}_h(x),
							\end{align}
							where the homogeneous term ${\Phi}_h$ satisfies
							\begin{align*}
								-\alpha {\Phi}_h^{\prime \prime}+ {\Phi}_h= 0,\ \  \Phi_h(0)= 0.
							\end{align*}
							It then follows that
							\begin{align}\label{phix}
								{\Phi}_h(x)=\sinh (\gamma x) A,
							\end{align}
							where $A \in \mathbb{R}^N$ is  an unknown constant vector to be determined by the right boundary condition in \eqref{equation-A}. Specifically, at $x=1$, we have
							\begin{subequations}\label{phip}
								\begin{align}
									\Phi(1) & =\Phi_p(1)+\sinh (\gamma) A, \\
									\Phi^{\prime}(1) & =\Phi^{\prime}_p(1)+\gamma \cosh (\gamma) A .
								\end{align}
							\end{subequations}
							By the definition of
							$\Phi_p$ (see \eqref{phipde}), it follows that
							\begin{subequations}\label{phip2}
								\begin{align}
									\Phi_p(1)=&-\int_0^1 \frac{\sinh (\gamma( 1-s)}{\alpha \gamma} \Gamma(s) \text{d}s,\\
									\Phi_p^{\prime}(1)=&-\int_0^1 \frac{\cosh (\gamma(1-s))}{\alpha} \Gamma(s)\text{d} s.
								\end{align}
							\end{subequations}
							Substituting \eqref{phip} and \eqref{phip2} into \eqref{20c}   yields
							\begin{align*}
								(\gamma \cosh (\gamma) I-\sinh (\gamma) B) A=B \Phi_{p}(1)-\Phi_p^{\prime}(1).
							\end{align*}
							Let $M:=\gamma \cosh (\gamma) I-\sinh (\gamma) B$.  A direct computation   yields
							\begin{align*}
								\det (M)=
								\sinh^N(\gamma) \left( \left( \gamma \coth(\gamma) + l \right)^N - \prod_{i=1}^N k_i \right).
							\end{align*}
							Since $k_i\in (0,l)$, we   have $ \prod_{i=1}^N k_i<  l^N$. Moreover, noting that $\gamma \coth(\gamma)>0$ and $\sinh^N(\gamma)>0$ for $\gamma > 0$, it follows that $ \left(  \gamma \coth(\gamma)  + l \right)^N>\prod_{i=1}^N k_i $. Therefore,  $\det (M)>0  $, which guarantees that the matrix $M$ is invertible. It then  follows that
							\begin{align}\label{A}
								A=M^{-1}\left(B \Phi_{p}(1)-\Phi_p^{\prime}(1)\right).
							\end{align}
							Hence, substituting \eqref{phipde}, \eqref{phix},  and \eqref{A} into \eqref{phi-ex}, we obtain
							\begin{align*}
								\Phi(x)=&-\int_0^x \frac{\sinh (\gamma(x-s))}{\alpha \gamma} \Gamma(s) \text{d} s+\sinh (\gamma x) M^{-1}\left(B \Phi_p(1)-\Phi_p^{\prime}(1)\right).
							\end{align*}
							Therefore,  $  I - \mathcal{A}$ is surjective.
							Together with the dissipativity of $\mathcal{A}$, it follows from the semigroup theory \cite[pp. 184, 187, Theorem 1.2 and   1.5]{pazy1983} that $\mathcal{A}$ generates a $C_0$-semigroup on $X$. Consequently, for  any $\Phi_0 \in D(\mathcal{A})$,     equation \eqref{Phi} admits a  unique  solution  {$ \Phi\in C(\mathbb{R}_{\geq 0};{D(\mathcal{A}))}\cap C^1(\mathbb{R}_{>0};X) $.}
						\end{pf6}
						
						Based on Proposition \ref{matchala}, we now prove Theorem~\ref{theorem-wellposedness}.
						
						\begin{pf5}
							By virtue of the equivalence between \eqref{mathcalu} and \eqref{Phi}, we further obtain {that} the $\widetilde{\nu}$-system admits a    solution, thereby guaranteeing that
							the $\widetilde u^{(i)}$-system admits a    solution.
						\end{pf5}

						\subsection{Well-Posedness Analysis for the Closed-Loop System}
						We now investigate the well-posedness of the closed-loop system composed of \eqref{muti-agent}--\eqref{disturbances-observer} under the control laws given by \eqref{control lawv}, which   is stated as the following theorem.
						\begin{theorem}\label{closed-loop}
							The closed-loop $(u^{(i)},u^{\text{ref}},v^{(i)}, \hat q^{(i)})$-system composed of  \eqref{muti-agent}--\eqref{disturbances-observer} under the  control  laws given by \eqref{control lawv}  admits a unique solution 	$(u^{(i)},u^{\text{ref}},v^{(i)}, \hat q^{(i)}) \in C(\mathbb{R}_{\geq 0};(H^2(0,1))^4)\cap C^1(\mathbb{R}_{\geq 0};(L^2(0,1))^4)$.
						\end{theorem}
						\begin{pf6}  Note that the closed-loop system exhibits a cascaded structure with coupling occurring through the boundary conditions, we can conduct the well-posedness analysis  sequentially. Specifically, we establish the well-posedness of the $u^{\text{ref}}$-system~\eqref{reference_system}, the $u^{(i)}$-system~\eqref{muti-agent}, the $v^{(i)}$-system~\eqref{v(x,t)}, and the $\hat q^{(i)}$-system~\eqref{disturbances-observer} in a step-by-step manner. Similar to the proof of Theorem~\ref{theorem-wellposedness}, this proof also relies on the lifting technique and semigroup theory. We only sketch the main procedure.

							\textbf{Step 1:} Establish the well-posedness of the  $u^{\text{ref}}$-system~\eqref{reference_system} and	the $u^{(i)}$-system~\eqref{muti-agent} under the control laws  given by \eqref{control lawv}.  
							Let
							$h^{\text{ref}}:=r-l(x^2-x)r$ and $\phi^{\text{ref}}:=u^{\text{ref}}-h^{\text{ref}}$. 	
							Define the operator $\mathcal{A}_{\text{ref}}: D(\mathcal{A}_{\text{ref}})\subset L^{2}(0,1)\rightarrow L^{2}(0,1)$ by $\mathcal{A}_{\text{ref}}\psi:=\alpha\psi^{\prime\prime}$ with
							\begin{align*}
								D(\mathcal{A}_{\text{ref}}):=\left\{\psi\in H^2(0,1)~\big|~\psi (0)=0,\psi^{\prime}(1)=-l\psi(1)\right\}.
							\end{align*} Then, the $\phi^{\text{ref}}$-system can be written  as
								\begin{align}\label{phiref}
									\dot{\phi}^{\text{ref}}=&\mathcal{A}_{\text{ref}}\phi^{\text{ref}}+f^{\text{ref}}(t,{\phi}^{\text{ref}}), \ \
									\phi^{\text{ref}}(0)=  \phi^{\text{ref}}_0,
								\end{align}
							where $
							f^{\text{ref}}(t,\phi^{\text{ref}}):=-\lambda\phi^{\text{ref}}  - h^{\text{ref}}_t+\alpha h^{\text{ref}}_{xx}-\lambda h^{\text{ref}}$
							and $\phi^{\text{ref}}_0:=u^{\text{ref}}_0(x)-h^{\text{ref}}(x,0)$.

							By an argument analogous to the proof of Proposition \ref{matchala},  system~\eqref{phiref} admits a unique   solution    $ \phi^{\text{ref}}\in C(\mathbb{R}_{\geq 0};{D(\mathcal{A}_{\text{ref}}))}\cap C^1(\mathbb{R}_{>0};L^2(0,1)) $.         It then follows from the transformation $u^{\text{ref}} = \phi^{\text{ref}} + h^{\text{ref}}$ that the
							$u^{\text{ref}}$-system~\eqref{reference_system}  admits  a unique solution.   Moreover, by  Theorem~\ref{theorem-wellposedness} and \eqref{u-uref}, the $u^{(i)}$-system~\eqref{muti-agent} also admits a unique solution.
							
							\textbf{Step 2:} Establish the well-posedness of the  $v^{(i)}$-system~\eqref{v(x,t)}.  	
							Let
							\begin{align}\label{w=u-v}
								w^{(i)}(x, t):=u^{(i)}(x, t)-v^{(i)}(x, t).
							\end{align}
							By \eqref{muti-agent} and \eqref{v(x,t)}, we have
							\begin{subequations}\label{u-v}
								\begin{align}
									w^{(i)}_t(x, t)=&\alpha w_{x x}^{(i)}(x, t) - \lambda w^{(i)}(x, t) +f^{(i)}(x,t), (x,t)\in Q_\infty, \label{u-v1}\\
									w^{(i)}(0, t) =&q^{(i)}(t),t\in\mathbb{R}_{> 0},  \label{f-u-v} \\
									w^{(i)}_x(1, t) =&-l w^{(i)}(1, t) , t\in\mathbb{R}_{> 0},   \\
									w^{(i)}(x, 0) =&w^{(i)}_0(x):=u^{(i)}_0(x)-v^{(i)}_0(x), x\in(0,1).
								\end{align}	
							\end{subequations}
							Note that system~\eqref{u-v} differs from the $u^{\text{ref}}$-system~\eqref{reference_system} only by the presence of the additional inhomogeneous term $f^{(i)}$ in system~\eqref{u-v}.  By redefining  $	f^{\text{ref}}(t,\phi^{\text{ref}}):=f^{(i)}-\lambda\phi^{\text{ref}}  - h^{\text{ref}}_t+\alpha h^{\text{ref}}_{xx}-\lambda h^{\text{ref}}$, system~\eqref{u-v} can be written in the same abstract form as \eqref{phiref}.
							Consequently,   by the  similar argument  to the proof of Proposition \ref{matchala}, the $w^{(i)}$-system~\eqref{u-v} admits a unique solution.
							Then, using \eqref{w=u-v} together with  Theorem~\ref{theorem-wellposedness}, we get the well-posdness of the $v^{(i)}$-system~\eqref{v(x,t)}.
							
							\textbf{Step 3:} Establish the well-posedness of the	$\hat q^{(i)}$-system~\eqref{disturbances-observer}.
							Substituting \eqref{1e} and \eqref{3e} into \eqref{disturbances-observer}, and invoking the well-posedness of the $u^{(i)}$- and $v^{(i)}$- systems, we can  establish the well-posedness of the $\hat q^{(i)}$-system. Indeed, let
							\begin{align*}
								\hat Q:=&\begin{matrix}
									\left(\hat q^{(1)},\dots,\hat q^{(N)}\right)	\end{matrix}^{\top},\notag\\
								\hat H:=&\begin{matrix}
									\left(\hat h^{(1)},\dots,\hat h^{(N)}\right)	\end{matrix}^{\top}
								:=\left(x - \frac{l+1}{l+2}x^2\right)\begin{pmatrix}
									u^{(1)}_x(0,t)-v^{(1)}_x(0,t)  \\
									\vdots \\
									u^{(N)}_x(0,t)-v^{(N)}_x(0,t)
								\end{pmatrix},
							\end{align*}
							and $\hat\Phi:=\hat Q-\hat H$.
							Define the operator $\hat{\mathcal{A}}: D(\hat{\mathcal{A}}) \subset X\rightarrow X:=(L^2(0,1))^N$ by $\hat{\mathcal{A}}\Psi:=	\operatorname{diag}\left(   {\alpha  }(\psi^{(1)})^{\prime\prime}, \dots, {\alpha  } (\psi^{(N)})^{\prime\prime}   \right)$ with  \begin{align*}
								D(\hat{\mathcal{A}}):=&\big\{ {\Psi}:=\begin{matrix} ( \psi^{(1)},\dots, \psi^{(N)}) \end{matrix}^{\top} \in (H^2(0,1))^N~\big|~ \Psi^{\prime}  (0)=0, \Psi^\prime(1)=-l \Psi(1)\big\}.
							\end{align*}
							Then, the $\hat q^{(i)}$-system~\eqref{disturbances-observer} can be written as
							\begin{align}\label{hatq}
								\dot{\hat\Phi} = \hat{\mathcal{A}}\hat{\Phi}+\hat{\mathcal{F}}(t,\hat\Phi),\ \
								\hat\Phi(0)= \hat\Phi_0,
							\end{align}
							where $\hat{\mathcal{F}}(t,\hat\Phi):=-\lambda\hat\Phi  - \hat H_t+\alpha \hat H_{xx}-\lambda \hat H$ and $\hat\Phi_0:=\hat Q(x,0)-\hat H(x,0)$.
							
							Analogous  to the proof of Proposition \ref{matchala},	 system~\eqref{hatq}   admits a  unique   solution  $ \hat\Phi\in C(\mathbb{R}_{\geq 0};D(\hat{\mathcal{A}}))\cap C^1(\mathbb{R}_{>0};X) $.       Consequently, the $\hat q^{(i)}$-system~\eqref{disturbances-observer} also admits a unique solution.
							
							Combining the results of Steps 1--3, we complete the proof of Theorem~\ref{closed-loop}.
						\end{pf6}

						\section{Stability Analysis}\label{V}
						In this section, we first establish the boundedness for the estimation error $q^{(i)}(t)-\hat q^{(i)}(0,t)$  by constructing weighted Lyapunov functions. These functions also provide some weighted estimates that will be needed later in other proofs. Then, using the GLM,  the recursion technique, and the obtained weighted estimates, we derive   robustness estimates in the max-norm  for the track error $\widetilde u^{(i)}$-system~\eqref{widetildeu3},  the synchronization error $\widetilde u^{(i,j)}$-system~\eqref{widetildeu-i-j}, and  the overall closed-loop system, respectively.

						\subsection{{Boundedness}  Analysis for  the Disturbance Estimation}
						In this subsection, we prove the {boundedness of} the disturbance estimation error $q^{(i)}(t)-\hat q^{(i)}(0,t)$, i.e., Theorem~\ref{theorem-1}.
						
						\begin{pf1}
							The proof of the  estimate \eqref{q-hatw} is based on a stability result of an auxiliary system with only in-domain disturbances. Indeed, let
							\begin{align}\label{widetilde q}
								\widetilde q^{(i)}(x,t):=u^{(i)}(x,t)-v^{(i)}(x,t)-\hat q^{(i)}(x,t).
							\end{align}
							It follows from  \eqref{1e},  \eqref{3e}, and \eqref{4b} that
							\begin{align*}
								\widetilde q^{(i)}_x(0,t)=u^{(i)}_x(0,t)-v^{(i)}_x(0,t)-\hat q^{(i)}_x(0,t)=0.
							\end{align*}
							Then,   by \eqref{muti-agent}, \eqref{v(x,t)}, \eqref{disturbances-observer}, and \eqref{widetilde q}, we have
							\begin{subequations}\label{disturbances-error}
								\begin{align}		
									\widetilde{q}_t^{(i)}(x, t)=&\alpha \widetilde{q}^{(i)}_{xx}(x, t)-\lambda \widetilde{q}^{(i)}(x, t)
									+f^{(i)}(x,t), (x,t)\in Q_\infty, \\
									\widetilde{q}_x^{(i)}(0, t)=&0,t\in\mathbb{R}_{> 0},\label{15a}\\
									\widetilde{q}_x^{(i)}(1, t)=& -l \widetilde{q}^{(i)}(1, t),t\in\mathbb{R}_{> 0},\label{15b} \\
									\widetilde{q}^{(i)}(x, 0)=& \widetilde{q}_0^{(i)}(x) , x\in(0,1),
								\end{align}
							\end{subequations}
							where $\widetilde{q}_0^{(i)}(x):=u^{(i)}_0(x)-v^{(i)}_0(x)-\hat q^{(i)}_0(x)$ denotes the initial data.

							For notational simplicity, we omit the index $i$ in the proof. For  any $p>1$ and ${\sigma}\in (0, \lambda )$, consider the weighted Lyapunov function
							\begin{align}\label{Vly}
								V(t):=\int_0^1 (e^{\sigma t}\widetilde{q}) ^{2p} \text{d}x.
							\end{align}
							By integrating by parts and using \eqref{15a} and \eqref{15b},  we have
							\begin{align*}
								\dot{V}(t)
								=&  2p\left(\alpha (e^{\sigma t}\widetilde{q})^{2p-1}(e^{\sigma t}\widetilde{q})_x\big|^{x=1}_{x=0} -\alpha(2p-1)\int_0^1(e^{\sigma t}\widetilde{q})^{2(p-1)}(e^{\sigma t}\widetilde{q})_x^2\text{d}x-(\lambda-\sigma)V(t) + \int_0^1  (e^{\sigma t}\widetilde{q}) ^{2p-1} e^{\sigma t}f \text{d}x\right)\notag\\
								\leq & -2p (\lambda-\sigma) V(t) +2p  \int_0^1  (e^{\sigma t}\widetilde{q}) ^{2p-1}(e^{\sigma t}f)\text{d}x.
							\end{align*}
							By the Young's inequality with any  $\varepsilon\in(0,\lambda-\sigma)$ (see \cite[p.~ 706]{Evans2010}), we have
							\begin{align*}
								\int_0^1  (e^{\sigma t}\widetilde{q}) ^{2p-1}(e^{\sigma t}f)\text{d}x \leq &   \varepsilon  \int_0^1  (e^{\sigma t}\widetilde{q}) ^{2p}\text{d}x+ C_{p,\varepsilon}  \int_0^1  (e^{\sigma t}f)^{2p}\text{d}x,
							\end{align*}
							where $C_{p,\varepsilon}:=\frac{1}{2p}\left(\frac{2p-1}{2p\varepsilon}\right)^{2p-1}$. Therefore, we obtain
							\begin{align*}
								\dot{V}(t)
								\leq & -2p (\lambda-\sigma-\varepsilon) V(t)
								+2p  C_{p,\varepsilon}\int_0^1   (e^{\sigma t}f)^{2p}\text{d}x,
							\end{align*}
							which implies that
							\begin{align}
							V(t)
								\leq & e^{-2p (\lambda-\sigma-\varepsilon)t}V(0)+2p  C_{p,\varepsilon}{\int_0^t\!\!\int_0^1 e^{-2p (\lambda-\sigma-\varepsilon)(t-s)}}   (e^{\sigma t}f(x,s))^{2p}\text{d}x\text{d}s\notag\\
							\leq & e^{-2p (\lambda-\sigma-\varepsilon)t}V(0) +2p  C_{p,\varepsilon} \| e^{\sigma \cdot}f\|_{C(\overline{Q}_t)}^{2p}{\int_0^t e^{-2p (\lambda-\sigma-\varepsilon)(t-s)}}  \text{d}s.\label{q-2p}
							\end{align}
							Note that
							\begin{align*}
								\int_0^t e^{-2p (\lambda-\sigma-\varepsilon)(t-s)}  \text{d}s=\frac{1}{2p(\lambda-\sigma-\varepsilon)} (1- e^{-2p (\lambda-\sigma-\varepsilon)t})
								\leq \frac{1}{2p(\lambda-\sigma-\varepsilon)}.
							\end{align*}
							We infer from \eqref{q-2p} that
							\begin{align}
								\left\|e^{\sigma t}\widetilde{q}  (\cdot,t)\right\|_{L^{2p}(0,1)}
								\leq & \left( \frac{C_{p,\varepsilon}}{\lambda-\sigma-\varepsilon} \right)^{\frac{1}{2p}} \left\|e^{\sigma \cdot}f\right\|_{C(\overline{Q}_t)} + e^{- (\lambda-\sigma-\varepsilon)t} \left\| \widetilde{q}_0\right\|_{L^{2p}(0,1)}.\label{q-2p-norm}
							\end{align}
							Note that $\lim_{p\to +\infty}\left( \frac{C_{p,\varepsilon}}{\lambda-\sigma-\varepsilon} \right)^{\frac{1}{2p}} =\frac{1}{\varepsilon} $.  {
								Letting $p\to +\infty$   in \eqref{q-2p-norm} and noting the continuity of $\widetilde q$, we obtain}
							\begin{align}\label{sigmaequ.16}
								\left\|e^{\sigma t}\widetilde{q}  (\cdot,t)\right\|_{C([0,1])}
								\leq & e^{- (\lambda-\sigma-\varepsilon)t} \left\| \widetilde{q}_0\right\|_{C([0,1])}+ \frac{1}{\varepsilon} \left\|e^{\sigma \cdot}f\right\|_{C(\overline{Q}_t)}. 
							\end{align}
							For any $T\in\mathbb{R}_{>0}$,  letting $t\rightarrow T$ and then, multiplying both sides of \eqref{sigmaequ.16} by $e^{-\sigma T}$, lead to
							\begin{align}\label{equ.16}
								\!\!\! 	\left\|\widetilde{q}  (\cdot,T)\right\|_{C([0,1])}
								\leq&  e^{ -(\lambda -\varepsilon) T} \| \widetilde{q}_0\|_{C([0,1])}
								+ \frac{1}{\varepsilon} \|f\|_{C(\overline{Q}_T)}. 
							\end{align}
							By \eqref{1b}, \eqref{3b}, and \eqref{widetilde q}, we have
							\begin{align}
								\left|q (T)-\hat q (0,T)\right|
								=|u (0,T)-v (0,T)-\hat q (0,T)|
								= |\widetilde q (0,T)|
								\leq\|\widetilde q (\cdot,T)\|_{C([0,1])}, \forall T\in \mathbb{R}_{>0}.\label{eq-15}
							\end{align}
							By \eqref{equ.16}, \eqref{eq-15}, and letting $\varepsilon \to \lambda-\sigma$, we obtain \eqref{q-hatw}.  
						\end{pf1}

								\begin{rem}
									Note that the above proof remains valid in the case $\sigma=0$. Nonetheless, the weighted Lyapunov function $V(t)$ defined in \eqref{Vly} for $\sigma\neq 0$ is employed here to obtain the weighted estimate \eqref{sigmaequ.16}, which will be subsequently used in the proofs of   Theorem~\ref{track-error}, Theorem~\ref{original main result}, and  Theorem~\ref{theorem 4}.
								\end{rem}
								
								\subsection{Robustness Analysis for the  Tracking Error System}
								In this subsection, we prove the {robust stability of   the tracking error $\widetilde u^{(i)}$-system~\eqref{widetildeu3}}, i.e., Theorem~\ref{track-error}. It is worth noting that unlike   the $\widetilde q^{(i)}$-system~\eqref{disturbances-error},  the  $\widetilde u^{(i)}$-system~\eqref{widetildeu3}  not only  involves {$\widetilde q^{(i)}(0,t)$ and external disturbance  $d^{(i)}_0(t)$} in the Dirichlet boundary condition \eqref{eq-12b} but also introduces the term {$\widetilde u^{(i-1)}(1,t)$} and external disturbance {$d^{(i)}_1(t)$} in the Robin boundary condition \eqref{12c}, bringing difficulties to the stability assessment.  {For this reason,} we  employ the GLM, the recursion technique, and the weighted estimate \eqref{sigmaequ.16}  to prove  Theorem~\ref{track-error}.
								
								In the sequel, let $\mathcal{M}_i, \mathcal{N}, \mathcal{C}_i$, and $\widetilde{\mathcal{C}}_i$ be the same as in the statement of  Theorem~\ref{track-error}.
								
								\begin{pf2}
									We {proceed with} the proof in  {four} steps.

									\textbf{Step 1:} Define truncation functions. 
									For an  arbitrary constant $s>1$, let
									\begin{align*}
										g(\theta ):= \begin{cases}\theta^{s},  \theta  \geq 0 \\
											0,  \theta <0\end{cases} \text{and }\
										G(\theta ):=\int_0^\theta  g(\tau) \mathrm{d} \tau,
									\end{align*}
									which satisfy
									\begin{subequations}\label{properties}
										\begin{align}
											g(\theta )  \geq& 0, g^{\prime}(\theta ) \geq 0, G(\theta ) \geq 0,  \forall \theta  \in \mathbb{R},\label{gG1} \\
											g(\theta ) =&G(\theta )=0, \forall \theta \in \mathbb{R}_{\leq 0}.\label{gG2}
										\end{align}
									\end{subequations}
									
									\textbf{Step 2:} Derive an upper bound estimate for $\widetilde{u}^{(i)}$.
									For any $\sigma\in (0,\lambda)$, let
									\begin{align*}
										{\overline{u}^{(i)}(x, t):=} e^{\sigma t} \widetilde{u}^{(i)}(x, t), \overline{u}^{(i)}_0(x):=\widetilde{u}^{(i)}_0(x),
										\overline{d}^{(i)}_1(t):=  e^{\sigma t} {d}^{(i)}_1(t), \overline{d}^{(i)}_0(t)	:=	e^{{\sigma} t}d^{(i)}_0(t),
										\overline{f}^{(i)}(x, t):=  e^{\sigma t} {f}^{(i)}(x, t).
									\end{align*}
									In view of \eqref{widetildeu3},	for any $ T\in\mathbb{R}_{>0}$,   $\overline u^{(i)}$ satisfies
									\begin{align*}
										\overline{u}^{(i)}_t(x, t)=&\alpha\overline{u}^{(i)}_{x x}(x, t)-(\lambda-\sigma) \overline{u}^{(i)}(x, t) +\overline{f}^{(i)}(x, t),(x,t)\in Q_T, \\ 
										\overline{u}^{(i)}(0, t)=&e^{\sigma t}\widetilde q^{(i)}(0,t)+\overline d^{(i)}_0(t), t\in(0,T),\\
										\overline{u}^{(i)}_x(1, t)=& -l\overline{u}^{(i)}(1,t)+\overline{d}^{(i)}_1(t) +k_ie^{\sigma t}\widetilde u^{(i-1)}(1,t),t\in(0,T),\\
										\overline{u}^{(i)}(x,0)=&\overline{u}^{(i)}_0(x),x\in(0,1).
									\end{align*}
									%
									Let
									\begin{align*}
										\overline m^{(i)}:= \max& \left\{
										\left\|e^{\sigma \cdot}\widetilde{q}^{(i)}(0,\cdot)+\overline d_0^{(i)}\right\|_{C([0,T])}, \frac{1}{l}\left\|\overline{d}^{(i)}_1+k_ie^{\sigma \cdot  }\widetilde u^{(i-1)}(1,\cdot)\right\|_{C([0,T])},\frac{1}{\lambda-\sigma}\left\|\overline f^{(i)}\right\|_{C(\overline{Q}_T)}, \left\|\overline u^{(i)}_0\right\|_{C([0,1])}\right\}.
									\end{align*}
									Note that
									\begin{align}\label{v le hat k}
										\overline{u}^{(i)}(0,t)\leq 	\overline m^{(i)},\forall t\in(0,T).
									\end{align}
									Using the definitions of   $g$ and $G$ and applying integration by parts, for any $t\in (0,T)$, we obtain
									\begin{align}\label{widetildev}
										\frac{\text{d}}{\text{d} t} \int_0^1 G\left(\overline{u}^{(i)}-	\overline m^{(i)}\right) \text{d} x
										= I_1+I_2+I_3+I_4,
									\end{align}
									where
									\begin{align*}
										I_1:=& \alpha g\left(\overline{u}^{(i)}(1, t)-	\overline m^{(i)}\right) \Big(-l\overline{u}^{(i)}(1,t)+\overline{d}^{(i)}_1(t) +k_ie^{\sigma t}\widetilde u^{(i-1)}(1,t)\Big),\\
										I_2:=& - \alpha g\left(\overline{u}^{(i)}(0, t)-	\overline m^{(i)}\right)\overline{u}^{(i)}_x(0,t),\\
										I_3:=& -\int_0^1 \alpha g^{\prime}\left(\overline{u}^{(i)} -	\overline m^{(i)}\right) \left(\overline{u}_x^{(i)}\right)^2  \text{d} x, \\
										I_4:=& \int_0^1 g\left(\overline{u}^{(i)}-	\overline m^{(i)}\right)\left(-({\lambda-\sigma})\overline{u}^{(i)}+\overline f^{(i)}\right) \text{d} x .
									\end{align*}
									
									For $I_1$,   if $\overline{u}^{(i)}(1,t)\leq 	\overline m^{(i)}$ for some $t\in (0,T)$,  then by \eqref{gG2} we have $g\left(\overline{u}^{(i)}(1,t)-\overline m^{(i)}\right)={0}$ for this $t $ and hence, $I_1=0$. If $\overline{u}^{(i)}(1,t)> 	\overline m^{(i)}$ for some $t\in (0,T)$, then by  the definition of $	\overline m^{(i)}$ and {$l>0$}, we obtain
									\begin{align*}
										\overline{u}^{(i)}(1,t)\geq  \frac{1}{{ l }} \left|\overline{d}^{(i)}_1(t)+k_ie^{\sigma t}\widetilde u^{(i-1)}(1,t)\right|.
									\end{align*} Therefore, for this $t $, we have
									\begin{align*}
									 {l} \overline{u}^{(i)}(1,t)-\overline{d}^{(i)}_1(t)-k_ie^{\sigma t}\widetilde u^{(i-1)}(1,t)
										\geq\left|\overline{d}^{(i)}_1(t)+k_ie^{\sigma t}\widetilde u^{(i-1)}(1,t)\right|- \overline{d}^{(i)}_1(t)-k_ie^{\sigma t}\widetilde u^{(i-1)}(1,t)
										\geq 0.
									\end{align*}
									{Thus,} in view of \eqref{gG1} and $\alpha>0$, we always have $I_1\leq 0$ in $(0,T)$.
									
									For $I_2$, by~\eqref{gG2} and~\eqref{v le hat k},  we have $g\left(\overline{u}^{(i)}(0,t)-	\overline m^{(i)}\right)=0$ for all $ t\in (0,T)$; hence $I_2=0$ in $(0,T)$.
									
									For $I_3$, by \eqref{gG1} and $\alpha>0$, we get $I_3\leq 0$ in $(0,T)$.
									
									For $I_4$,   if $\overline{u}^{(i)}(x,t) \leq 	\overline m^{(i)}$ for some $(x,t)\in Q_T$,  then by  \eqref{gG2}, we obtain $g\left(\overline{u}^{(i)}(x,t) -	\overline m^{(i)}\right)=0$ for such $(x,t) $, and hence $I_4=0$ for the same $(x,t) $. If $\overline{u}^{(i)}(x,t)  > 	\overline m^{(i)}$  for some $(x,t)\in Q_T$, then by the definition of $	\overline m^{(i)}$ and $\lambda-\sigma> 0$, we have $\overline{u}^{(i)}(x,t) \geq \frac{1}{\lambda-\sigma}\left|\overline{f}^{(i)}(x,t) \right| $  for such $(x,t) $, which yields
									\begin{align*}
										(\lambda-\sigma) \overline{u}^{(i)}(x,t)  -\overline{f}^{(i)}(x,t) \geq\left|\overline{f}^{(i)}(x,t) \right| - \overline{f}^{(i)}(x,t)   \geq 0.
									\end{align*}
									In view of \eqref{gG1}, we always have $I_4\leq 0$  for all $(x,t)\in Q_T$.
									
									Since  $I_i\le 0$ for all $i=1,2,3,4$, it follows from \eqref{widetildev} that
									\begin{align*}
										\frac{\text{d}}{\text{d} t} \int_0^1 G\left(\overline{u}^{(i)}-	\overline m^{(i)}\right) \text{d} x \leq 0,\forall t\in (0,T).
									\end{align*}
									Therefore,  for any $t\in (0,T)$, we have
									\begin{align*}
										\int_0^1 G\left(\overline{u}^{(i)} -	\overline m^{(i)}\right) \text{d} x \leq \int_0^1 G\left(\overline{u}^{(i)}_0 -	\overline m^{(i)}\right) \text{d} x\leq 0,
									\end{align*}
									which, along with $\overline{u}^{(i)}_0 \leq	\overline m^{(i)} $, the property of $G$ (see \eqref{gG2}), and the continuity of $\overline{u}^{(i)}$, implies that
									\begin{align*}
										\overline{u}^{(i)}(x,t)\leq 	\overline m^{(i)},{\forall (x,t)\in  Q_T}.
									\end{align*}
									Then, for any $t\in (0,T)$, we have
									\begin{align}\label{upper}
										e^{\sigma t} \widetilde{u}^{(i)}(x,t)
										\leq&   \left\|\overline u^{(i)}_0\right\|_{C([0,1])}+\frac{1}{\lambda-\sigma}\left\|\overline f^{(i)}\right\|_{C(\overline{Q}_T)} +\frac{1}{l}\left\|\overline{d}^{(i)}_1+k_ie^{\sigma \cdot}\widetilde u^{(i-1)}(1,\cdot)\right\|_{C([0,T])}\notag\\
										&+\left\|e^{\sigma \cdot}\widetilde{q}^{(i)}(0,\cdot)+\overline d_0^{(i)}\right\|_{C([0,T])}.
									\end{align}

									\textbf{Step 3:} Derive a lower bound estimate for $\widetilde{u}^{(i)}$. Note that equation \eqref{widetildeu3} is linear. Considering  $-\widetilde{u}^{(i)}$
									and proceeding in the same way as in Step 2, we obtain for any $t\in (0,T)$
												\begin{align}\label{lower}
														 -e^{{\sigma} t} \widetilde{u}^{(i)}(x,t)
													\leq &
													\left\| { \widetilde u^{(i)}_0}\right\|_{C([0,1])}+\frac{1}{\lambda-\sigma}\left\| {e^{\sigma \cdot} f^{(i)}}\right\|_{C(\overline{Q}_T)}  +\frac{1}{l}\left\| {e^{\sigma \cdot} {d}^{(i)}_1}+k_ie^{\sigma \cdot}\widetilde u^{(i-1)}(1,\cdot)\right\|_{C([0,T])}\notag\\
															&+\left\|e^{\sigma \cdot}\widetilde{q}^{(i)}(0,\cdot)+ {e^{\sigma \cdot} d_0^{(i)}}\right\|_{C([0,T])},
												\end{align}
												where  $\sigma\in (0, \lambda )$ is the same as Step 2.
												
												\textbf{Step 4:} Derive the   estimate for $\widetilde{u}^{(i)}$. By ~\eqref{upper} and \eqref{lower}, for any $ t\in (0,T)$,  we have
												\begin{align}\label{defenition psi}
													\left\|e^{\sigma t}\widetilde{u}^{(i)}(\cdot,t)\right\|_{C([0,1])}
													\leq&\left\|\widetilde{u}^{(i)}_0\right\|_{C([0,1])}+\frac{1}{\lambda-\sigma}\left\|e^{\sigma \cdot}{f}^{(i)}\right\|_{C\left(\overline Q_T\right)}+\frac{1}{{l}}\left\|e^{\sigma \cdot}{d}^{(i)}_1+k_ie^{\sigma \cdot}\widetilde u^{(i-1)}(1,\cdot)\right\|_{C([0,T])}\notag\\
													&+\left\|e^{\sigma \cdot}\widetilde{q}^{(i)}(0,\cdot)\right\|_{C([0,T])}+\left\|e^{\sigma \cdot} d_0^{(i)}\right\|_{C([0,T])}.
												\end{align}
												Letting $\varepsilon \to \lambda-\sigma$ in \eqref{sigmaequ.16}, it holds that for any $t\in (0,T)$
												\begin{align}\label{eq-25}
													\!\!\!  e^{\sigma t}  \left| \widetilde{q}^{(i)}(0,t)\right|
													\leq &  \left\| \widetilde q_0^{(i)}\right\|_{C([0,\!1])} \!\! +\!\frac{{ 1} }{\lambda-\!\sigma}\left\|e^{\sigma \cdot}f^{(i)}\right\|_{C(\overline{Q}_t)}.
												\end{align}
												Note that
												\begin{align}\label{f}
													\left\|e^{\sigma \cdot}f^{(i)}\right\|_{C(\overline{Q}_t)}\leq \left\|e^{\sigma \cdot}f^{(i)}\right\|_{C(\overline{Q}_T)},\forall t\in (0,T).
												\end{align}
												Substituting \eqref{eq-25} and \eqref{f} into \eqref{defenition psi},  for any $t\in (0,T)$, we obtain
												\begin{align}\label{eq-24'}
												\left\|e^{\sigma t}\widetilde{u}^{(i)}(\cdot,t)\right\|_{C([0,1])}
													\leq& \left\|\widetilde{u}^{(i)}_0\right\|_{C([0,1])}+\left\|\widetilde q^{(i)}_0\right\|_{C([0,1])}+ {\frac{2}{\lambda-\sigma}\left\|e^{\sigma \cdot}{f}^{(i)}\right\|_{C\left(\overline Q_T\right)}}+\frac{1}{{l}}\left\|e^{\sigma \cdot}{d}^{(i)}_1\right\|_{C([0,T])}\notag\\
													&+ \left\|e^{\sigma \cdot}d_0^{(i)}\right\|_{{C([0,T])}} +\frac{k_i}{{l}}\left\|e^{\sigma \cdot}\widetilde u^{(i-1)}(1,\cdot)\right\|_{C([0,T])}.
												\end{align}
												From \eqref{eq-24'},  at the point $x=1$, we deduce that for any $t\in(0,T)$
												\begin{align}\label{eq24'}
													\left|e^{\sigma t}\widetilde{u}^{(i)}(1,t)\right|
													\leq& \left\|\widetilde{u}^{(i)}_0\right\|_{C([0,1])}+\left\|\widetilde q^{(i)}_0\right\|_{C([0,1])}+ {\frac{2}{\lambda-\sigma}\left\|e^{\sigma \cdot}{f}^{(i)}\right\|_{C\left(\overline Q_T\right)}}+\frac{1}{{l}}\left\|e^{\sigma \cdot}{d}^{(i)}_1\right\|_{C([0,T])}\notag\\
													&+ \left\|e^{\sigma \cdot}d_0^{(i)}\right\|_{{C([0,T])}} +\frac{k_i}{{l}}\left\|e^{\sigma \cdot}\widetilde u^{(i-1)}(1,\cdot)\right\|_{C([0,T])}.
												\end{align}
												Then,   applying  the continuity of $\widetilde{u}^{(i)}$ in $t$ and the arbitrariness of $t\in(0,T) $ to \eqref{eq24'}, we obtain
												\begin{align}\label{eq-24}
													\left\|e^{\sigma \cdot}\widetilde{u}^{(i)}(1,\cdot) \right\|_{C([0,T])}
													\leq& \left\|\widetilde{u}^{(i)}_0\right\|_{C([0,1])}+\left\|\widetilde q^{(i)}_0\right\|_{C([0,1])}+\frac{2}{\lambda-\sigma}\left\|e^{\sigma \cdot}{f}^{(i)}\right\|_{C\left(\overline Q_T\right)}+\frac{1}{{l}}\left\|e^{\sigma \cdot}{d}^{(i)}_1\right\|_{C([0,T])}\notag\\
													&+ \left\|e^{\sigma \cdot}d_0^{(i)}\right\|_{{C([0,T])}} +\frac{k_i}{{l}}\left\|e^{\sigma \cdot}\widetilde u^{(i-1)}(1,\cdot)\right\|_{C([0,T])}.
												\end{align}
												Now, using    {recursion} for \eqref{eq-24}, we have
												\begin{align}\label{qi-1}
													\left\|e^{\sigma \cdot}\widetilde{u}^{(i)}(1,\cdot)\right\|_{C([0,T])}
													\leq& \left(1+\frac{k_i}{{l}}\right)\left( \max_{i} \left\|\widetilde{u}^{(i)}_0\right\|_{C([0,1])}\right.\left.+\max_{i}\left\|\widetilde q^{(i)}_0\right\|_{C([0,1])}\right) +\left(1+\frac{k_i}{l}\right) \mathcal{E}(T)\notag\\
													& +\frac{k_ik_{i-1}}{{l^2}}\left\|e^{\sigma\cdot  }\widetilde u^{(i-2)}(1,\cdot)\right\|_{C([0,T])}\notag\\
													\leq& \cdots\notag\\
													\leq&\mathcal{M}_i\left( \max_{i} \left\|\widetilde{u}^{(i)}_0\right\|_{C([0,1])}\right.\left.+\max_{i}\left\|\widetilde q^{(i)}_0\right\|_{C([0,1])}\right) +\mathcal{M}_i\mathcal{E}(T) +\mathcal{N}\left\|e^{\sigma\cdot  }\widetilde u^{(i)}(1,\cdot)\right\|_{C([0,T])},
												\end{align}
												where
												\begin{align}\label{E(T)}
													\mathcal{E}(T):= 	  \frac{2}{\lambda-\sigma}\max_{i}\left\|e^{\sigma \cdot}{f}^{(i)}\right\|_{C(\overline{Q}_T)}\!
													+\frac{1}{l}\max_{i} \!\left\|e^{\sigma \cdot}{d^{(i)}_1}\right\|_{C([0,T])} 	+ \max_{i}\left\|e^{\sigma \cdot}{d^{(i)}_0}\right\|_{C([0,T])}.
												\end{align}
												Noting that $\mathcal{N}<1$ and rearranging terms  for \eqref{qi-1},   we  obtain
												\begin{align}\label{eq-26-}
												 \left\|e^{\sigma \cdot}\widetilde u^{(i)}(1,\cdot)\right\|_{C([0,T])} 	\leq {\mathcal{C}_i\mathcal{E}(T)}+\mathcal{C}_i\!\left( \!\max_{i} \left\|\widetilde{u}^{(i)}_0\right\|_{C([0,1])} \!+\max_{i}\left\|\widetilde q^{(i)}_0\right\|_{C([0,1])}\right).
												\end{align}	
												Putting \eqref{eq-26-} into \eqref{eq-24'}, and then letting $t\to T$, we get
													\begin{align}\label{eq-26}
													\left\|e^{\sigma T}\widetilde u^{(i)}( \cdot,T)\right\|_{C([0,1])}	\leq \widetilde{\mathcal{C}}_i\mathcal{E}(T)+\widetilde{\mathcal{C}}_i  \left( \max_{i} \left\|\widetilde{u}^{(i)}_0\right\|_{C([0,1])} +\max_{i}\left\|\widetilde q^{(i)}_0\right\|_{C([0,1])}\right),
													\end{align}
													which {implies  \eqref{tildeu-es}}.
												\end{pf2}									
												\subsection{Robustness Analysis for the  Synchronization Error System}
												In this subsection, we prove the {robust stability of} the synchronization error
												$\widetilde u^{(i,j)}$-system~\eqref{widetildeu-i-j}, i.e., Theorem~\ref{original main result}.
												
												In the sequel, let $\sigma,\mathcal{M}_i, \mathcal{N}, \mathcal{C}_i, \widetilde{\mathcal{C}}_i$, $\mathcal{H}_{i,j}$ and $\widetilde{\mathcal{H}}_{i,j}$ be the same as in the statement of  Theorem~\ref{track-error} and Theorem~\ref{original main result}.
												
												\begin{pf3}
													The proof is also based   on the application of the GLM, the recursion technique, and the weighted estimate \eqref{sigmaequ.16}, and can {proceed} in a similar way to the proof of Theorem~\ref{track-error}. {Therefore, we} only {outline} the {main differences}.
													
													Indeed, taking  $d^{(i,j)}_1(t)+k_i\widetilde u^{(i-1)}(1,t)-k_j\widetilde u^{j-1}(1,t)$ as an entire disturbance,  we find that the synchronization error
													$\widetilde u^{(i,j)}$-system~\eqref{widetildeu-i-j} has a similar form to the  {tracking  error $\widetilde{u}^{(i)}$-system~\eqref{widetildeu3}}. Note that
													\begin{align*}
													 k_i\widetilde u^{(i-1)}(1,t)-k_j\widetilde u^{j-1}(1,t)
														= (k_i-k_j)\widetilde u^{(i-1)}(1,t)+k_j\widetilde u^{(i-1,j-1)}(1,t).
													\end{align*}
													Analogous to the proof of \eqref{eq-24'} for the $\widetilde u^{(i)}$-system~\eqref{widetildeu3}, for any $t \in (0,T)$ with an arbitrary $T\in \mathbb{R}_{>0}$, we have
													\begin{align}\label{eq48}
												  \left\|e^{\sigma t}\widetilde{u}^{(i,j)}(\cdot,t)\right\|_{C([0,1])}
													 \leq& \left\|\widetilde{u}^{(i,j)}_0\right\|_{C([0,1])}+\left\|\widetilde q^{(i,j)}_0\right\|_{C([0,1])}+\frac{2}{\lambda-\sigma}\left\|e^{\sigma \cdot}{f}^{(i,j)}\right\|_{C\left(\overline Q_T\right)}+\frac{1}{{l}}\left\|e^{\sigma \cdot}{d}^{(i,j)}_1\right\|_{C([0,T])}\notag\\
													 &+ \left\|e^{\sigma \cdot}d_0^{(i,j)}\right\|_{C([0,T])} +\frac{|k_i-k_j|}{{l}}\left\|e^{\sigma \cdot}\widetilde u^{(i-1)}(1,\cdot)\right\|_{C([0,T])}\notag\\
													 &+\frac{k_j}{{l}}\left\|e^{\sigma \cdot}\widetilde u^{(i-1,j-1)}(1,\cdot)\right\|_{C([0,T])}.
													\end{align}
													In particular, at the point $x=1$,  applying  the continuity of $\widetilde{u}^{(i,j)}$ in $t$  and   the arbitrariness of $t\in(0,T) $  to \eqref{eq48}, we obtain	
													\begin{align}\label{36}
													 \left\|e^{\sigma \cdot}\widetilde{u}^{(i,j)}(1,\cdot)\right\|_{C([0,T])}
													\leq& \left\|\widetilde{u}^{(i,j)}_0\right\|_{C([0,1])}+\left\|\widetilde q^{(i,j)}_0\right\|_{C([0,1])}+\frac{2}{\lambda-\sigma}\left\|e^{\sigma \cdot}{f}^{(i,j)}\right\|_{C\left(\overline Q_T\right)}+\frac{1}{{l}}\left\|e^{\sigma \cdot}{d}^{(i,j)}_1\right\|_{C([0,T])}\notag\\
													&+ \left\|e^{\sigma \cdot}d_0^{(i,j)}\right\|_{C([0,T])}  +\frac{|k_i-k_j|}{{l}}\left\|e^{\sigma \cdot}\widetilde u^{(i-1)}(1,\cdot)\right\|_{C([0,T])} \notag\\
													&+\frac{k_j}{{l}}\left\|e^{\sigma \cdot}\widetilde u^{(i-1,j-1)}(1,\cdot)\right\|_{C([0,T])}.
													\end{align}
													Using  recursion  for \eqref{36}, we deduce that
													\begin{align}\label{eq49}
													  \left\|e^{\sigma \cdot}\widetilde{u}^{(i,j)}(1,\cdot)\right\|_{C([0,T])}
													 \leq&\bigg(1+\frac{k_j}{l}\bigg)\bigg(\max_{i,j}\left\|\widetilde{u}^{(i,j)}_0\right\|_{C([0,1])}+\max_{i,j}\left\|\widetilde q^{(i,j)}_0\right\|_{C([0,1])}\bigg)+\bigg(1+\frac{k_j}{l}\bigg)\widetilde{\mathcal{E}}(T)\notag\\
														&    +\bigg(\frac{|k_i-\!k_j|}{l}+\!\frac{k_j|k_{i-1}-\!k_{j-1}|}{l^2}\bigg)\!\max_{i}\!\left\|e^{\sigma \cdot}\widetilde u^{(i)}(1,\!\cdot)\right\|_{C([0,T])} \notag\\
														&+\frac{k_jk_{j-1}}{l^2}\left\|e^{\sigma \cdot}\widetilde u^{(i-2,j-2)}(1,\cdot)\right\|_{C([0,T])} \notag\\
													 	\leq& \cdots\notag\\
													 \leq& \mathcal{M}_j\bigg(\max_{i,j}\left\|\widetilde{u}^{(i,j)}_0\right\|_{C([0,1])}+\max_{i,j}\left\|\widetilde q^{(i,j)}_0\right\|_{C([0,1])}\bigg)+\mathcal{H}_{i,j}\max_i\left\|e^{\sigma \cdot} \widetilde u^{(i)}(1,\cdot)\right\|_{C([0,T])} \notag\\
													 &+\mathcal{M}_{j}\widetilde{\mathcal{E}}(T)+\mathcal{N}\left\|e^{\sigma \cdot}\widetilde u^{(i,j)}(1,\cdot)\right\|_{C([0,T])},
													\end{align}
												where
												\begin{align*}
													\widetilde{\mathcal{E}}(T):=& \frac{2}{\lambda-\sigma}\max_{i,j}\left\|e^{\sigma \cdot}{f}^{(i,j)}\right\|_{C\left(\overline Q_T\right)}  +\frac{1}{{l}}\max_{i,j}\left\|e^{\sigma \cdot}{d}^{(i,j)}_1\right\|_{C([0,T])}  + \max_{i,j}\left\|e^{\sigma \cdot}d_0^{(i,j)}\right\|_{C([0,T])}.
												\end{align*}
												Analogous to  \eqref{eq-26-}, noting  that $\mathcal{N}<1$ and rearranging terms  for \eqref{eq49},   we  obtain 	
												\begin{align}\label{eq50}
												   \left\|e^{\sigma \cdot}\widetilde{u}^{(i,j)}(1,\cdot)\right\|_{C([0,T])}
													\leq & {\mathcal{C}}_j\bigg(\max_{i,j}\left\|\widetilde{u}^{(i,j)}_0\right\|_{C([0,1])}+\max_{i,j}\left\|\widetilde q^{(i,j)}_0\right\|_{C([0,1])}\bigg)+{\mathcal{C}}_j\widetilde{\mathcal{E}}(T)\notag\\
													&+\frac{{\mathcal{H}}_{i,j}}{1-\mathcal{N}}\max_i\left\|e^{\sigma \cdot} \widetilde u^{(i)}(1,\cdot)\right\|_{C([0,T])}.
												\end{align}
												Putting \eqref{eq50} into \eqref{eq48}, and then letting $t\rightarrow T$, we have
												\begin{align*}
												 \left\|e^{\sigma T}\widetilde{u}^{(i,j)}(\cdot,T)\right\|_{C([0,1])}
													\leq& \widetilde{\mathcal{C}}_j\bigg(\max_{i,j}\left\|\widetilde{u}^{(i,j)}_0\right\|_{C([0,1])}+\max_{i,j}\left\|\widetilde q^{(i,j)}_0\right\|_{C([0,1])}\bigg)+\widetilde{\mathcal{C}}_j\widetilde{\mathcal{E}}(T)\notag\\
													&+\widetilde{\mathcal{H}}_{i,j}\max_i\left\|e^{\sigma \cdot} \widetilde u^{(i)}(1,\cdot)\right\|_{C([0,T])},
												\end{align*}
												which, along with   \eqref{tildeu-es}, implies \eqref{tildeu-i-j-es}.  
											\end{pf3}
											\subsection{ISS Analysis for the  Closed-Loop System}
											
											In this subsection,  we prove  the ISS property of the closed-loop system composed of \eqref{muti-agent}--\eqref{disturbances-observer} under the control  laws  given by \eqref{control lawv}, i.e., Theorem \ref{theorem 4}. 
											Specifically, by using the GLM, we first sequentially derive weighted   estimates  for $ \hat{q}^{(i)} $, $   v^{(i)} $, $ u^{\text{ref}} $, and $ u^{(i)}  $. Then, using these estimates, we establish the ISS in the max-norm  for the closed-loop system.
											
											In the sequel, let $\sigma, \mathcal{C}_i$,  $\widetilde{\mathcal{C}}_i$, and $\mathcal{E}(T)$ be the same as in the statement of Theorem~\ref{track-error} and \eqref{E(T)}.
											
											\begin{pf4}
												The proof is divided into four steps.
												
												\textbf{Step 1:} Establish the estimate for $\left\| e^{\sigma t} \hat{q}^{(i)}(\cdot, t)\right\|_{C([0,1])}$. From \eqref{w=u-v} and \eqref{widetilde q}, we obtain, we obtain
												\begin{align}\label{w-hatq}
													\hat q^{(i)}(x,t)	=w^{(i)}(x,t)-\widetilde q^{(i)}(x,t).
												\end{align}
												For any $t\in (0,T)$ with an arbitrary $T\in\mathbb{R}_{>0}$, multiplying both sides of \eqref{w-hatq} by $e^{\sigma t}$   and
												applying the triangle inequality yield
												\begin{align}\label{hat q}
													\left\| e^{\sigma t} \hat{q}^{(i)}(\cdot, t) \right\|_{C([0,1])}
													\le     \left\| e^{\sigma t} w^{(i)}(\cdot, t) \right\|_{C([0,1])}
													 + \left\| e^{\sigma t} \widetilde{q}^{(i)}(\cdot, t) \right\|_{C([0,1])}.
												\end{align}
												
												We need to  estimate the two terms on the right-hand side of \eqref{hat q}.  Indeed, for the first term,   	similar  to the proof of Theorem~\ref{track-error}, applying the GLM to \eqref{u-v}, we obtain
												\begin{align}\label{w}
												 	\left\|e^{\sigma t} w^{(i)}(\cdot, t)\right\|_{C([0,1])}
													\leq   \left\|w_0^{(i)}\right\|_{C([0,1])} +\frac{1}{\lambda\!-\sigma}\left\|e^{\sigma \cdot} f^{(i)}\right\|_{C\left(\overline{Q}_T\right)} +\left\|e^{\sigma \cdot }q^{(i)}\right\|_{C([0,T])},  \forall t \in (0,T).
												\end{align}
												For the second term,		letting $\varepsilon \to \lambda-\sigma$ in \eqref{sigmaequ.16}, it holds that
												\begin{align}\label{widetildeq}
													\left\|e^{\sigma t}\widetilde{q}^{(i)}  (\cdot,t)\right\|_{C([0,1])}
													\leq    \frac{1}{\lambda-\sigma} \left\|e^{\sigma \cdot}f^{(i)}\right\|_{C(\overline{Q}_t)}  +\left\| \widetilde{q}^{(i)}_0\right\|_{C([0,1])},\forall t\in(0,T).
												\end{align}
												
												Then, noting that  $t\in(0,T)$ is arbitrary, and substituting \eqref{f}, \eqref{w}, and \eqref{widetildeq} into \eqref{hat q}, we have
												\begin{align}\label{hat qsigma}
												 	\left\| e^{\sigma t} \hat{q}^{(i)}(\cdot, t)\right\|_{C([0,1])}
													\leq   \left\|w_0^{(i)}\right\|_{C([0,1])}+\left\| \widetilde q_0^{(i)}\right\|_{C([0,1])}+\left\|e^{\sigma \cdot }q^{(i)}\right\|_{C([0,T])}  +\frac{2}{\lambda-\sigma}\left\|e^{\sigma \cdot} f^{(i)}\right\|_{C\left(\overline{Q}_T\right)} ,\forall t\in(0,T).
												\end{align}
												Furthermore, since     $t\in(0,T)$ is arbitrary and   $\hat{q}^{(i)}$ is continuous in $t$, we obtain that at  the point $x=0$
												\begin{align}\label{hat q 0}
												 	\left\| e^{\sigma \cdot} \hat{q}^{(i)}(0, \cdot) \right\|_{C([0,T])}
													\le  \left\|w_0^{(i)}\right\|_{C([0,1])}+\left\| \widetilde q_0^{(i)}\right\|_{C([0,\!1])} +\frac{2}{\lambda-\sigma}\left\|e^{\sigma \cdot} f^{(i)}\right\|_{C\left(\overline{Q}_T\right)} +\left\|e^{\sigma \cdot }q^{(i)}\right\|_{C([0,T])}.
												\end{align}
												\textbf{Step 2:} Establish the estimate for $\left\| e^{\sigma t} v^{(i)}(\cdot, t)\right\|_{C([0,1])}$. Under the control laws given by \eqref{control lawv}, the $v^{(i)}$-system~\eqref{v(x,t)} is written as
												\begin{align*}
													v_t^{(i)}(x,t) &= \alpha  v_{xx}^{(i)}(x,t) - \lambda v^{(i)}(x,t),(x,t)\in Q_\infty,  \\
													v^{(i)}(0,t) &= r(t) - \hat{q}^{(i)}(0,t) + d_0^{(i)}(t), t\in\mathbb{R}_{> 0},\\
													v_x^{(i)}(1,t) &= -l v^{(i)}(1,t) + k_i \widetilde{u}^{(i-1)}(1,t) + d_1^{(i)}(t), t\in\mathbb{R}_{> 0},\\
													v^{(i)}(x, 0)&=v_0^{(i)}(x), x\in(0,1).
												\end{align*}
												As in the proof of Theorem~\ref{track-error}, using the GLM, we obtain
												\begin{align}\label{esigmav}
												 \left\| e^{\sigma t} v^{(i)}(\cdot, t) \right\|_{C([0,1])}
													\le    & \left\| v_0^{(i)} \right\|_{C([0,1])} + \left\| e^{\sigma \cdot} r \right\|_{C([0,T])} + \left\| e^{\sigma \cdot} \hat{q}(0, \cdot) \right\|_{C([0,T])} + \left\| e^{\sigma \cdot} d_0^{(i)} \right\|_{C([0,T])} \notag\\
													& + \frac{k_i}{l} \left\| e^{\sigma \cdot} \widetilde{u}^{(i-1)}(1, \cdot) \right\|_{C([0,T])}  + \frac{1}{l} \left\| e^{\sigma \cdot} d_1^{(i)} \right\|_{C([0,T])},\forall t\in(0,T).
												\end{align}
												
												Then, putting \eqref{hat q 0} and \eqref{eq-26-} into \eqref{esigmav}, we get
												\begin{align}\label{esigmatv}
												 \left\| e^{\sigma t} v^{(i)}(\cdot, t) \right\|_{C([0,1])}
													\le&  \left\| v_0^{(i)} \right\|_{C([0,1])} + \left\| w_0^{(i)} \right\|_{C([0,1])} + \left\| \widetilde{q}_0^{(i)} \right\|_{C([0,1])}    + \left\| e^{\sigma \cdot} r \right\|_{C([0,T])} + \left\| e^{\sigma \cdot} q^{(i)} \right\|_{C([0,T])}  \notag\\
													&  + \left\| e^{\sigma \cdot} d_0^{(i)} \right\|_{C([0,T])}+ \frac{1}{l} \left\| e^{\sigma \cdot} d_1^{(i)} \right\|_{C([0,T])} + \frac{k_i}{l} \left\| e^{\sigma \cdot} \widetilde{u}^{(i-1)}(1, \cdot) \right\|_{C([0,T])}  \notag\\
													& + \frac{2}{\lambda-\sigma} \left\| e^{\sigma \cdot} f^{(i)} \right\|_{C(\overline{Q}_T)}\notag\\
													\le &  \left\| v_0^{(i)} \right\|_{C([0,1])}\! + \!\left\| w_0^{(i)} \right\|_{C([0,1])}  +  \frac{k_i \mathcal{C}_{i-1}}{l} \max_{i} \left\|\widetilde{u}^{(i)}_0\right\|_{C([0,1])} + \left( 1 + \frac{k_i \mathcal{C}_{i-1}}{l} \right) \max_{i}\left\|\widetilde q^{(i)}_0\right\|_{C([0,1])} \notag\\
													& + \| e^{\sigma \cdot} r \|_{C([0,T])} + \left\| e^{\sigma \cdot} q^{(i)} \right\|_{C([0,T])} + \left\| e^{\sigma \cdot} d_0^{(i)} \right\|_{C([0,T])}  + \frac{1}{l} \left\| e^{\sigma \cdot} d_1^{(i)} \right\|_{C([0,T])}\notag\\
													& + \frac{2}{\lambda-\sigma} \left\| e^{\sigma \cdot} f^{(i)} \right\|_{C(\overline{Q}_T)} + \frac{k_i \mathcal{C}_{i-1}}{l} \mathcal{E}(T),\forall t\in(0,T).
												\end{align}
												\textbf{Step 3:}  Establish the estimates for $\left\| e^{\sigma t} u^{\text{ref}}(\cdot, t)\right\|_{C([0,1])}$ and $\left\| e^{\sigma t} u^{(i)}(\cdot, t)\right\|_{C([0,1])}$.
												As in the proof of Theorem~\ref{track-error}, applying the GLM to the reference system~\eqref{reference_system}, we obtain
												\begin{align}\label{esigmauref}
												\left\| e^{\sigma t} u^{\text{ref}}(\cdot, t) \right\|_{C([0,1])}
													\le  \left\| u_0^{\text{ref}} \right\|_{C([0,1])} + \left\| e^{\sigma \cdot} r \right\|_{C([0,T])},\forall t\in(0,T).
												\end{align}
												Using the definition of $\widetilde{u}^{(i)}$ (see \eqref{u-uref}) and  the triangle inequality, we have for all $t\in(0,T)$
												\begin{align}\label{esigmau}
													\left\| e^{\sigma t} u^{(i)}(\cdot, t) \right\|_{C([0,1])}
													\le \left\| e^{\sigma t} \widetilde{u}^{(i)}(\cdot, t) \right\|_{C([0,1])} + \left\| e^{\sigma t} u^{\text{ref}}(\cdot, t) \right\|_{C([0,1])}.
												\end{align}
												Letting $T:=t$ in \eqref{eq-26}       and substituting it and \eqref{esigmauref} into \eqref{esigmau},  we obtain  for any $t\in(0,T)$				
												\begin{align}\label{esigmauend}
													\left\| e^{\sigma t} u^{(i)}(\cdot, t) \right\|_{C([0,1])}
													\leq &\left\| u_0^{\text{ref}} \right\|_{C([0,1])} + \| e^{\sigma \cdot} r \|_{C([0,T])}+{\widetilde{\mathcal{C}}_i\mathcal{E}(T)}+\widetilde{\mathcal{C}}_i \! \left( \max_{i} \left\|\widetilde{u}^{(i)}_0\right\|_{C([0,1])} \!+\max_{i}\left\|\widetilde q^{(i)}_0\right\|_{C([0,1])}\right).
												\end{align}
												
												\textbf{Step 4:} Conclution. Combining with \eqref{hat qsigma}, \eqref{esigmatv}, \eqref{esigmauref}, and \eqref{esigmauend}, we get for any $t\in(0,T)$
												\begin{align}\label{esigmat}
												& \left\| e^{\sigma t} u^{(i)}(\cdot, t) \right\|_{C([0,1])} + \left\| e^{\sigma t} u^{\text{ref}}(\cdot, t) \right\|_{C([0,1])} + \left\| e^{\sigma t} \hat{q}^{(i)}(\cdot, t) \right\|_{C([0,1])} + \left\| e^{\sigma t} v^{(i)}(\cdot, t) \right\|_{C([0,1])} \notag\\
													\le\ & 2\left\| u_0^{\text{ref}} \right\|_{C([0,1])} + 2\left\| w_0^{(i)} \right\|_{C([0,1])} + \left\| v_0^{(i)} \right\|_{C([0,1])}+ \left\| \widetilde q_0^{(i)}\right\|_{C([0,1])} + 3\left\| e^{\sigma \cdot} r \right\|_{C([0,T])} + 2\left\| e^{\sigma \cdot} q^{(i)} \right\|_{C([0,T])}  \notag\\
													&+ \frac{4}{\lambda-\sigma} \left\| e^{\sigma \cdot} f^{(i)} \right\|_{C(\overline{Q}_T)} + \left\| e^{\sigma \cdot} d_0^{(i)} \right\|_{C([0,T])} + \frac{1}{l} \left\| e^{\sigma \cdot} d_1^{(i)} \right\|_{C([0,T])}  + \left( \frac{k_i \mathcal{C}_{i-1}}{l} + \widetilde{\mathcal{C}}_i \right) \notag\\
													&\times\left( \max_{i} \left\|\widetilde{u}^{(i)}_0\right\|_{C([0,1])} + \mathcal{E}(T) \right)  + \left( 1 + \frac{k_i \mathcal{C}_{i-1}}{l} + \widetilde{\mathcal{C}}_i \right) \max_{i}\left\|\widetilde q^{(i)}_0\right\|_{C([0,1])}.
												\end{align}
												Applying the triangle inequality to \eqref{u-uref}, \eqref{w=u-v},  and \eqref{widetilde q}, we obtain
												\begin{align}
													\left\|\widetilde{u}^{(i)}_0\right\|_{C([0,1])}\leq& \left\| {u}^{(i)}_0\right\|_{C([0,1])}+\left\| {u}^{\text{ref}}_0\right\|_{C([0,1])},\label{trangle2}\\
													\left\| w_0^{(i)}\right\|_{C([0,1])}\leq& \left\| u_0^{(i)}\right\|_{C([0,1])}+\left\| v_0^{(i)}\right\|_{C([0,1])},\label{trangle1}\\
													\left\| \widetilde q_0^{(i)}\right\|_{C([0,1])}\leq& \left\| u_0^{(i)}\right\|_{C([0,1])}+\left\| v_0^{(i)}\right\|_{C([0,1])}+\left\| \hat q_0^{(i)}\right\|_{C([0,1])}.\label{trangle3}
												\end{align}
												Then, substituting \eqref{trangle2}, \eqref{trangle1}, and \eqref{trangle3} into \eqref{esigmat}, and  letting $t\to T$, we get
												\begin{align*}
													& \left\| e^{\sigma T} u^{(i)}(\cdot, T) \right\|_{C([0,1])} + \left\| e^{\sigma T} u^{\text{ref}}(\cdot, T) \right\|_{C([0,1])} + \left\| e^{\sigma T} \hat{q}^{(i)}(\cdot, T) \right\|_{C([0,1])} + \left\| e^{\sigma T} v^{(i)}(\cdot, T) \right\|_{C([0,1])} \notag\\
													\le & \left( 1  +2 \widetilde{\mathcal{C}}_i \right) \left\| u_0^{\text{ref}} \right\|_{C([0,1])}  + 2\left( 1  +2 \widetilde{\mathcal{C}}_i \right) \max_{i} \left\| u_0^{(i)} \right\|_{C([0,1])} + \left( 4  + 2\widetilde{\mathcal{C}}_i \right) \left( \max_{i} \left\| v_0^{(i)} \right\|_{C([0,1])} + \max_{i} \left\| \hat q_0^{(i)} \right\|_{C([0,1])} \right) \notag\\
													& + \left( 1 +2 \widetilde{\mathcal{C}}_i \right) \mathcal{E}(T) + 3\left\| e^{\sigma \cdot} r \right\|_{C([0,T])} + 2\left\| e^{\sigma \cdot} q^{(i)} \right\|_{C([0,T])},
												\end{align*}
												which implies  \eqref{closedloop}.
											\end{pf4}

											\begin{rem}
												Note that the estimates \eqref{w}, \eqref{widetildeq}, and \eqref{esigmauref} imply  that the $w^{(i)}$-system~\eqref{u-v} is ISS w.r.t. $f^{(i)}$ and $q^{(i)}$, the $\widetilde q^{(i)}$-system~\eqref{disturbances-error} is ISS w.r.t. $f^{(i)}$, and the $u^{\mathrm{ref}}$-system~\eqref{reference_system} is ISS w.r.t. $r$, respectively. 
											\end{rem}

										\section{Numerical Results}\label{Sec. VI}
										In this section, we perform numerical experiments to illustrate the effectiveness of the proposed control scheme and to demonstrate the {boundedness} of the estimation error $|q^{(i)}(t)-\hat q^{(i)}(0,t)|$,  the tracking error   $\widetilde u^{(i)}$-system    \eqref{widetildeu3}, and the synchronization error $  u^{(i,j)}$-system~\eqref{widetildeu-i-j}. We also verify the  ISS  of the closed-loop system composed of \eqref{muti-agent}--\eqref{disturbances-observer} under the control laws  given by \eqref{control lawv}.

										In simulation, the parameters of the MAS~\eqref{muti-agent}  are selected as
											$ N=5,   \alpha=1,  \lambda=5,   l=1$.
												The control gains are chosen as
											$ k_1=0.1, k_2=0.2, k_3=0.3, k_4=0.4,  k_5=0.5$.
											
											The initial values of the    $u^{(i)}$-system~\eqref{muti-agent} and the $u^{\text{ref}}$-system~\eqref{reference_system} are set to
											\begin{align*}
												\begin{pmatrix}u^{(1)}_0(x)\\
													u^{(2)}_0(x)\\
													u^{(3)}_0(x)\\
													u^{(4)}_0(x)\\
													u^{(5)}_0(x)\end{pmatrix}
												= &\begin{pmatrix}-0.2\sin  (\pi(x-0.5) )\\
													5\sin (1.5\pi(x-0.5) )\\
													4\sin  (\pi(2x-0.5) )\\
													3.5\cos (\pi(x-1) )\\
													4\cos (\pi(2x-1) )\end{pmatrix}
											\end{align*}
											and
											\begin{align*}
												u^{\text{ref}}_0(x)=  2.8\sin(0.5\pi(x-0.5)),
											\end{align*}
											respectively.	
											
											The initial values of the {$\hat q^{(i)}$-system} \eqref{disturbances-observer}, the  $\widetilde q^{(i)}$-system~\eqref{disturbances-error}, and the $v^{(i)}$-system~\eqref{v(x,t)}  are set to  
											\begin{align*}
												\hat q^{(i)}_0(x)= 0,
												\begin{pmatrix}\widetilde q^{(1)}_0(x)\\
													\widetilde q^{(2)}_0(x)\\
													\widetilde q^{(3)}_0(x)\\
													\widetilde q^{(4)}_0(x)\\
													\widetilde q^{(5)}_0(x) \end{pmatrix} = \begin{pmatrix} -1.6\sin (\pi(x-0.5) )\\
													2\sin (1.5\pi(x-0.5) )\\
													2\sin (\pi(2x-0.5) )\\
													1.6 \cos(2\pi x+1.2) )\\
													1.2 \cos(\pi(2x-1) ) \end{pmatrix},
											\end{align*}
											and
											\begin{align*}
												v^{(i)}_0(x)=u^{(i)}_0(x)-\widetilde q^{(i)}_0(x) ,
											\end{align*}
											respectively.

											The reference signal and  observable   boundary disturbances are chosen as
											\begin{align*}	
												r=D_0\sin(10t)
											\end{align*}	
											and
											\begin{align*}	
												\begin{pmatrix}
													q^{(1)}(t)\\
													q^{(2)}(t)\\
													q^{(3)}(t)\\
													q^{(4)}(t)\\
													q^{(5)}(t) \end{pmatrix} =D_1\begin{pmatrix} \sin(2t)\\0.5\cos(5t)\\0.1t + \sin(t)\\
													2 - e^{-t}\\
													0.5\sin(2t+1)\end{pmatrix},\end{align*}
											respectively, where $D_0,D_1\in\{0,1,3,5\}$ are used to characterize the  amplitudes of the signal inputs and disturbances.
											
											The  unobservable   in-domain disturbances are taken as 	
											\begin{align*}\begin{pmatrix}
													f^{(1)}(x,t)\\
													f^{(2)}(x,t)\\
													f^{(3)}(x,t)\\
													f^{(4)}(x,t)\\
													f^{(5)}(x,t) \end{pmatrix} =A\begin{pmatrix} -1 + \sin(x + 10 t)\\  1.2 + \cos(x + 10 t)\\
													0.8 + \sin(x+10 t)\\ -1 + \sin(2 x + 10 t)\\
													1 + \cos(2 x + 10 t) \end{pmatrix}.
											\end{align*}
											
											The   unobservable Dirichlet and Robin boundary disturbances are set to  	\begin{align*}
												\begin{pmatrix}
													d_{0}^{(1)}(t)\\
													d_{0}^{(2)}(t)\\
													d_{0}^{(3)}(t)\\
													d_{0}^{(4)}(t)\\
													d_{0}^{(5)}(t)
												\end{pmatrix}
												= A_0\begin{pmatrix} \sin(5+10t)\\
													\cos(2+10t)\\
													\sin(1+10t)\\
													\cos(2+10t)\\
													\sin(4+10t)\end{pmatrix}  \end{align*}    and  \begin{align*}\begin{pmatrix}
													d_{1}^{(1)}\\ d_{1}^{(2)}\\ d_{1}^{(3)}\\d_{1}^{(4)}\\ d_{1}^{(5)} \end{pmatrix} =A_1\begin{pmatrix}\sin(1+10t)\\
													\sin(2+10t)\\
													\cos(10t)\\
													\sin(10t)\\
													\sin(1+10t)\end{pmatrix}, \end{align*}
											respectively, where $A\in\{0,2,4\}$ and $A_0,A_1\in\{0,3,5\}$ are used to characterize the  amplitudes of disturbances.
											
\begin{figure}[htbp]
												\begin{center}									\includegraphics[scale= 0.3]{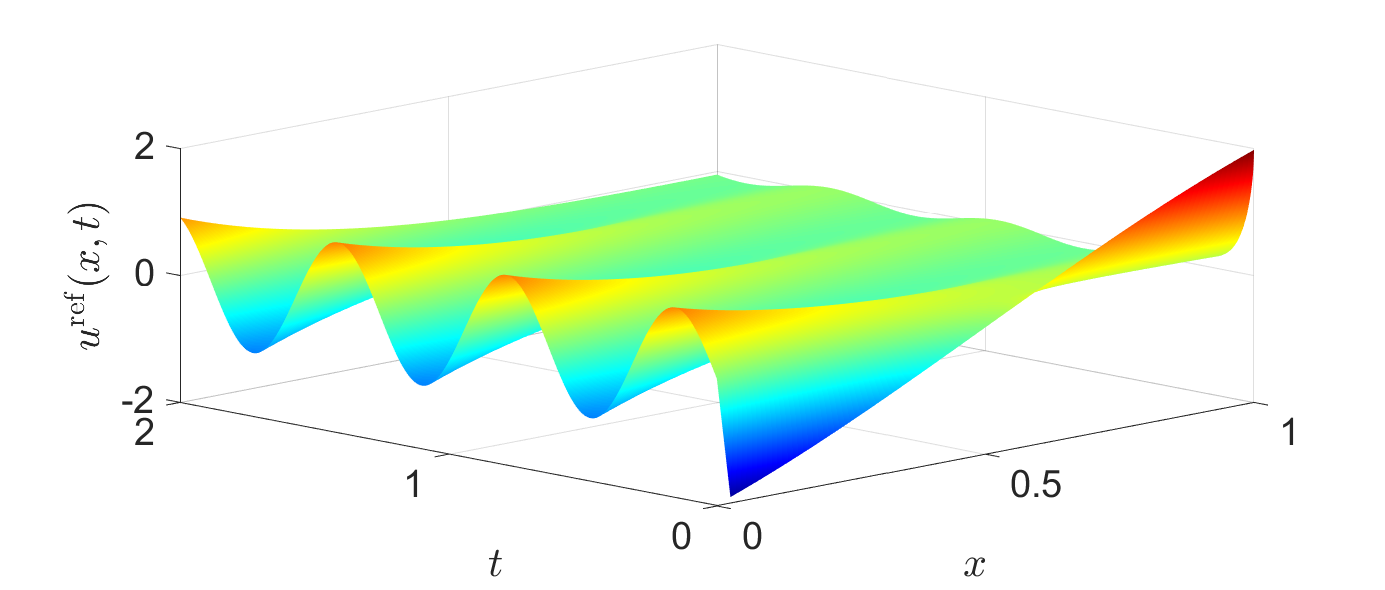}\\
													\includegraphics[scale= 0.3]{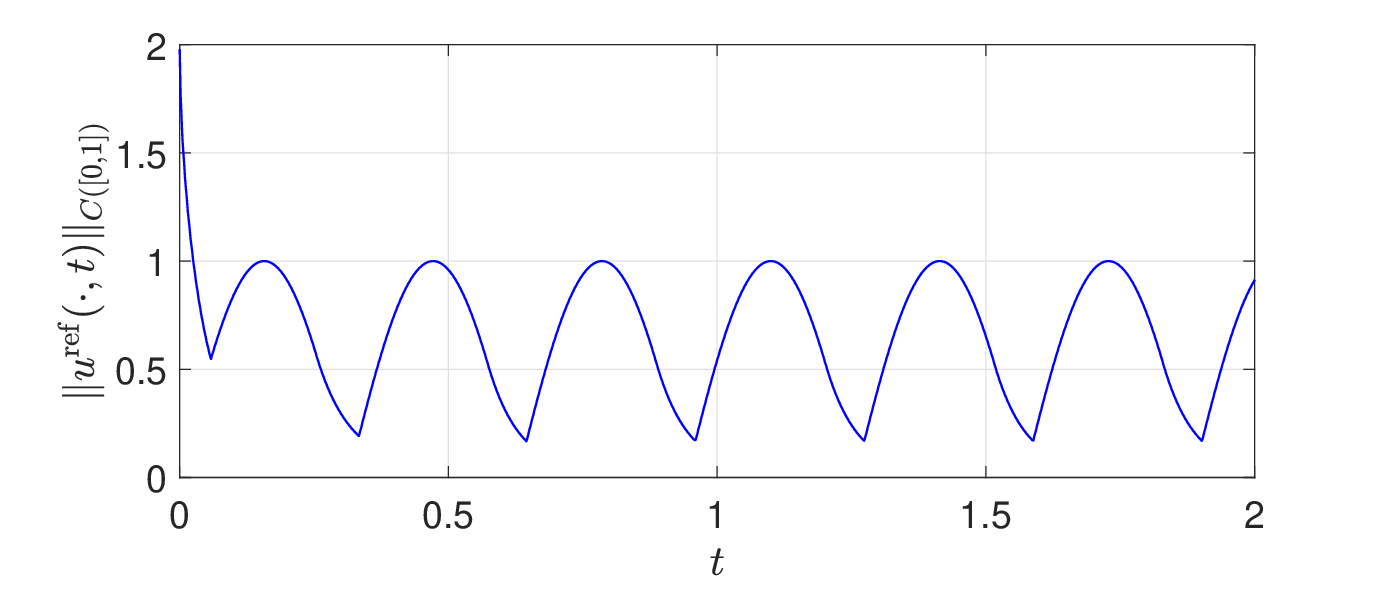}
													\caption{Evolution of   $u^{\text{ref}}(x,t)$ (top) and  $\|u^{\text{ref}}(\cdot,t)\|_{C([0,1])}$ (bottom) for  the reference system~\eqref{reference_system} when $D_0=1$.}
													\label{fig1}
												\end{center}
											\end{figure}

											Figure \ref{fig1} shows  the evolution of the  solution and its norm in the max-norm  for the reference  $u^{\text{ref}}$-system~\eqref{reference_system} with $D_0=1$, whose state   exhibits   oscillations and remains bounded.
											
\begin{figure}[htbp]
												\begin{center}
													\includegraphics[scale= 0.3]{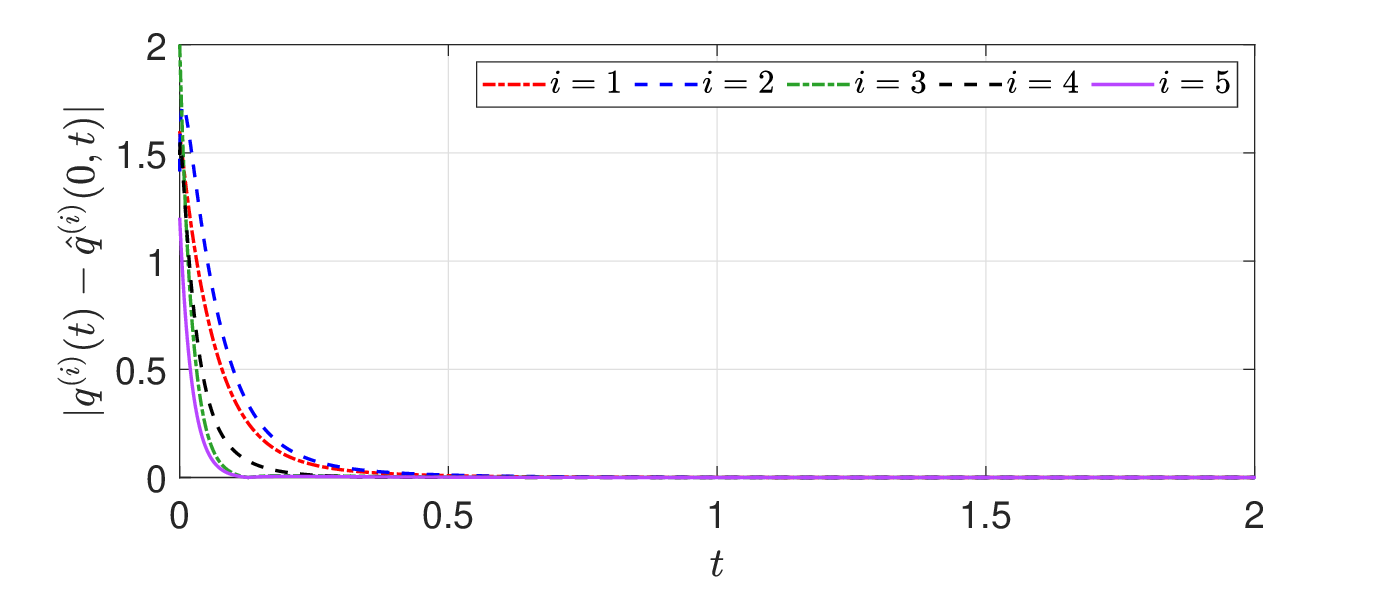}\\
													\includegraphics[scale= 0.3]{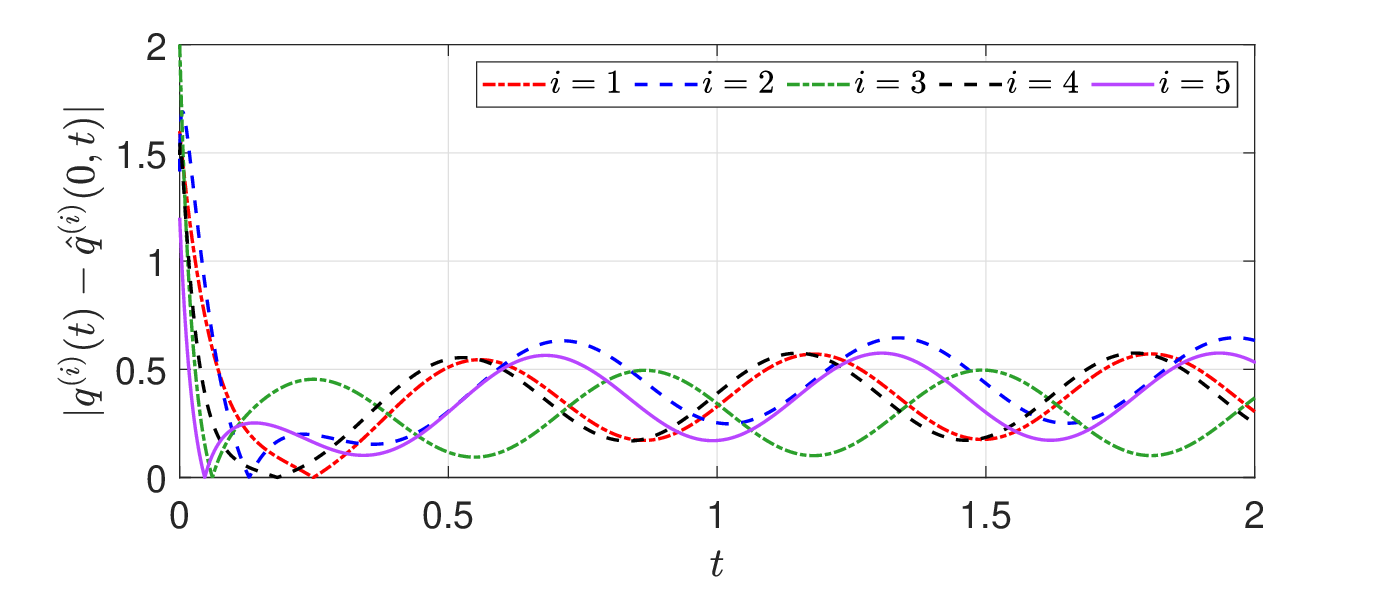}\\
													\includegraphics[scale= 0.3]{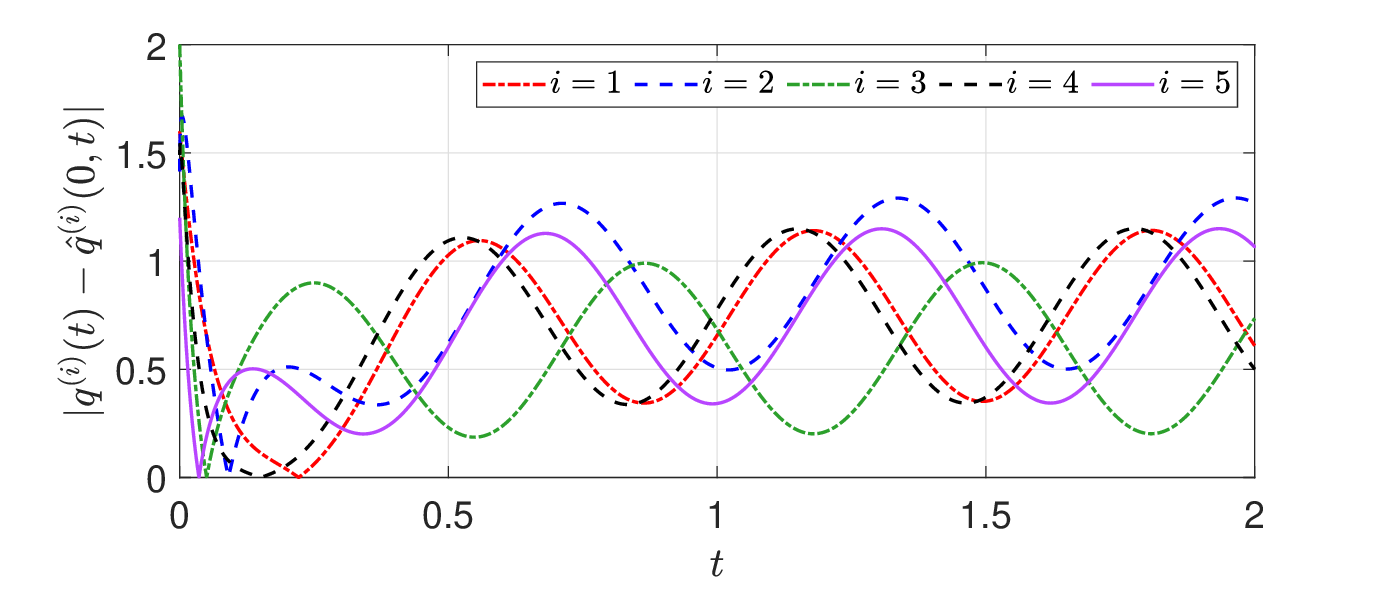}
													\caption{Evolution of the estimation error $|q^{(i)}(t)-\hat q^{(i)}(0,t)|$ with different in-domain disturbance $f^{(i)}(x,t)$ when $D_1=1$:  $A=0$ (top),   $A=2$ (middle), and $A=4$ (bottom).}
													\label{fig2}
												\end{center}
											\end{figure}

											Figure \ref{fig2}   shows the evolution of the error between the   Dirichlet boundary  disturbances $q^{(i)}  $ with $D_1=1$ and their estimation $\hat q^{(i)} (0,t)$. In the absence of unobservable disturbances $f^{(i)},d^{(i)}_0 $, and $d^{(i)}_1$, it can be seen that the Dirichlet boundary  disturbance estimation error  for each agent  is relatively large at the initial stage due to the incomplete convergence of the observer state.
											With increasing time, the observer gradually captures the dynamic characteristics of the disturbances $q^{(i)}$, and the estimation error   rapidly decays    to a small neighborhood of zero. In the presence of  unobservable disturbances $f^{(i)}$, it can be seen that the estimation error  remains bounded. Moreover, the  bottom and middle subplots of Fig. \ref{fig2}  show  that, as the amplitudes of the in-domain disturbance  decrease, the oscillation amplitudes of the estimation error decrease accordingly.   Fig. \ref{fig2}  well illustrates the {boundedness} of the estimation error   $|q^{(i)}(t)-\hat q^{(i)}(0,t)|$ in the presence of unobservable disturbances $f^{(i)}$.

\begin{figure}[htbp]
												\centering
												\includegraphics[scale= 0.3]{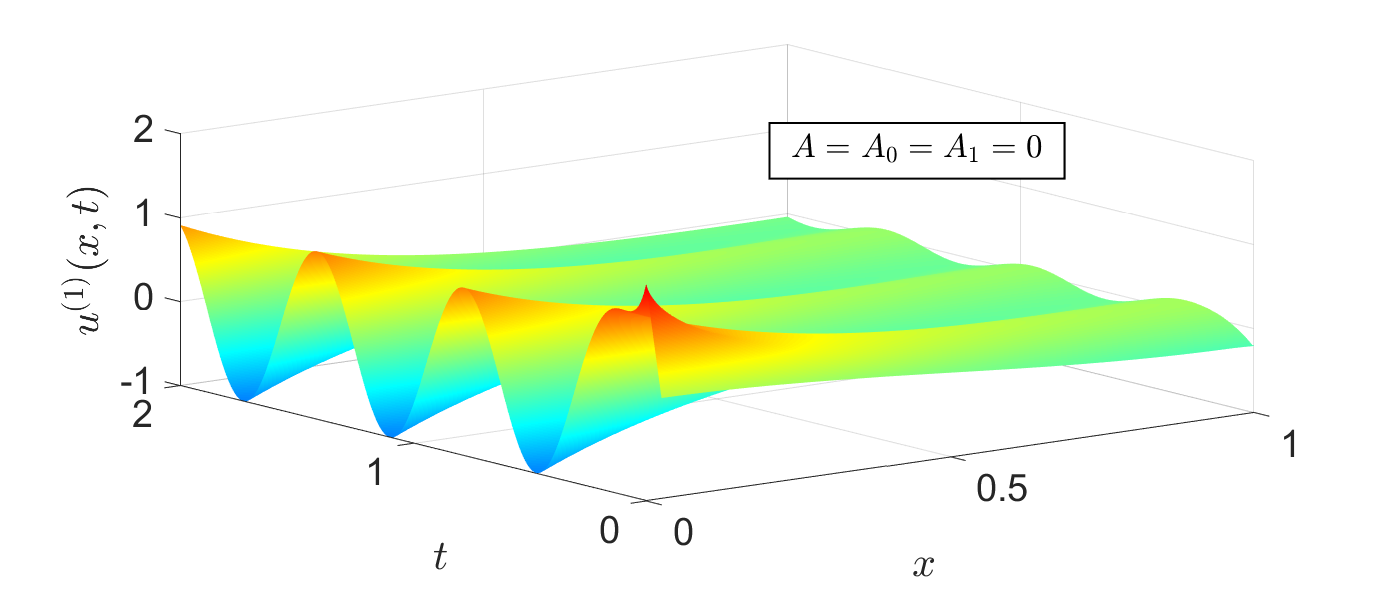}
												\includegraphics[scale= 0.3]{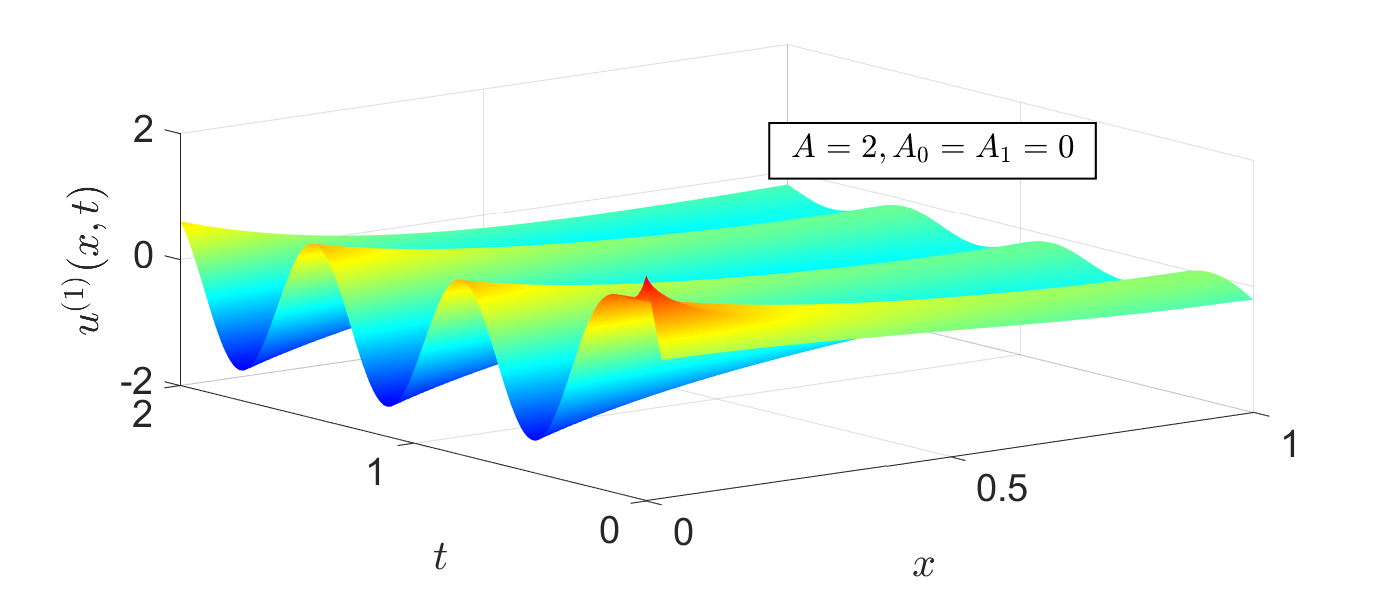}
												\includegraphics[scale= 0.3]{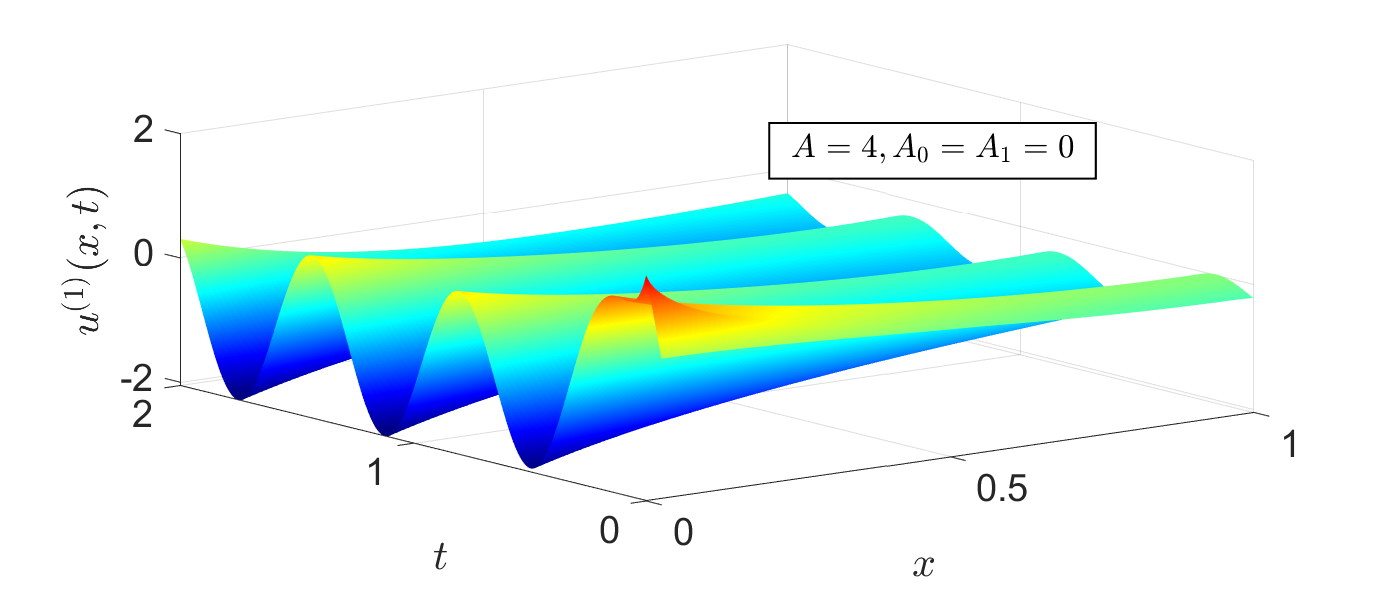}
												\includegraphics[scale= 0.3]{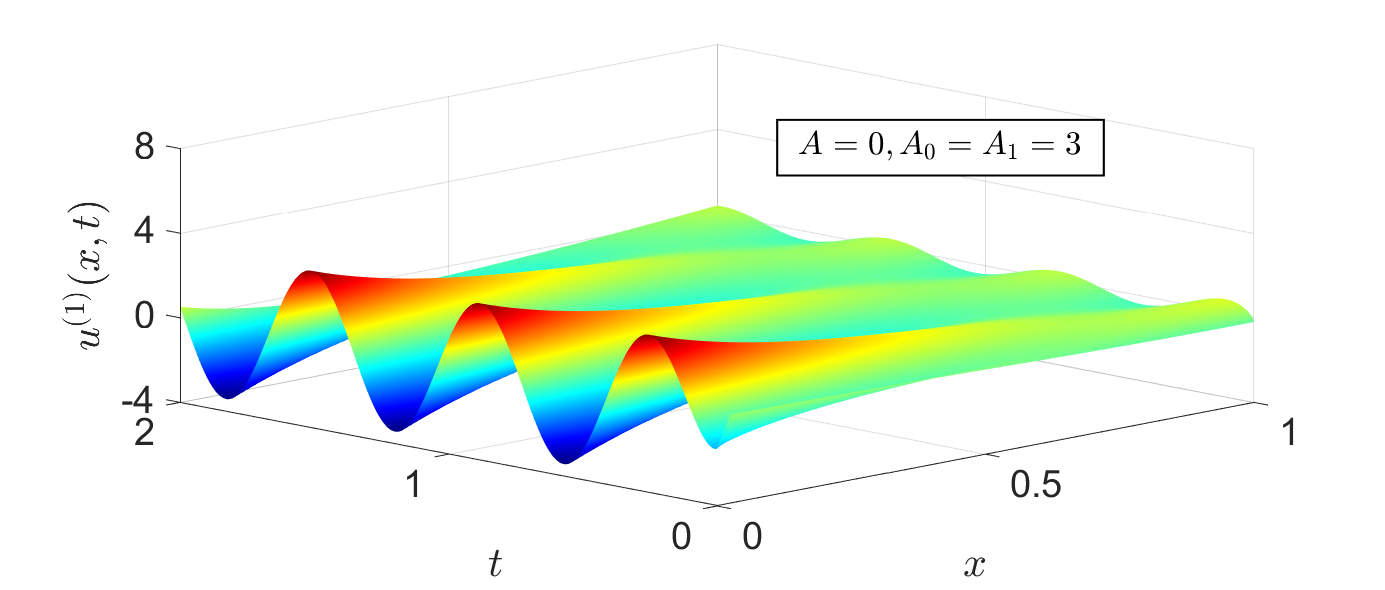}
												\includegraphics[scale= 0.3]{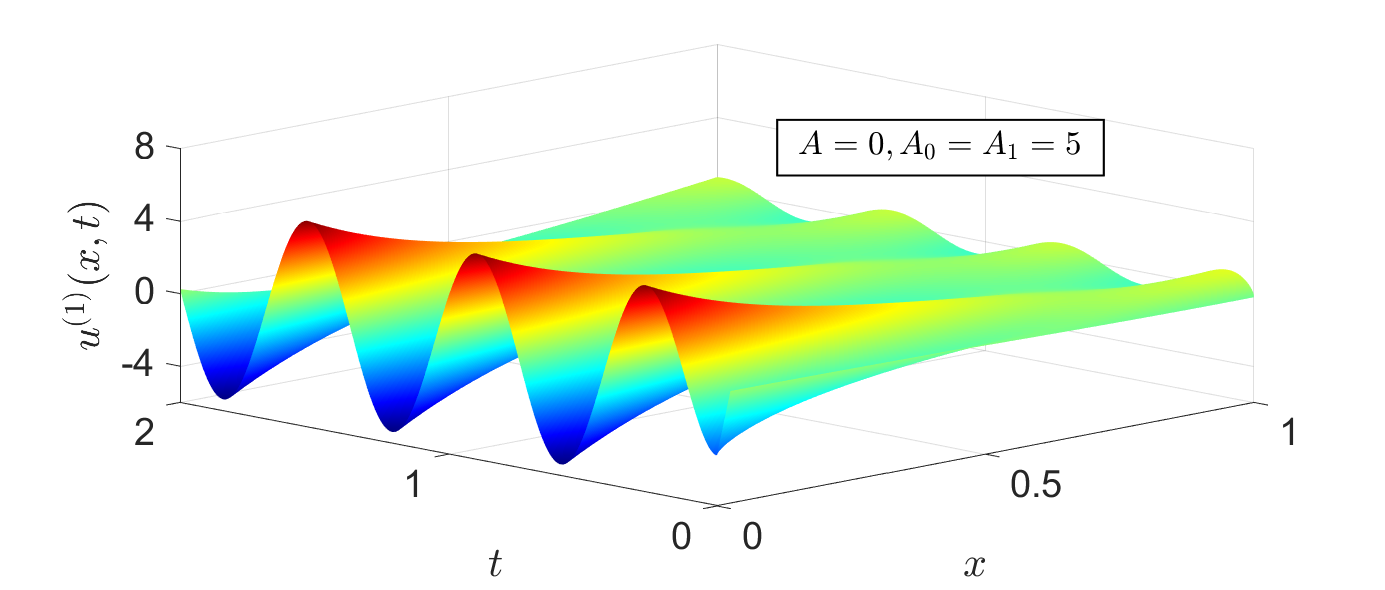}
												\caption{Evolution  of  $u^{(1)}(x,t)$   for the first agent in system \eqref{muti-agent}   with  different  $f^{(i)}(x,t)$, $d^{(i)}_0(t)$, and $ d^{(i)}_1(t)$ when $D_0=D_1=1$.}
												\label{fig5}
											\end{figure}

\begin{figure}[htbp]
												\begin{center}
													\includegraphics[scale= 0.3]{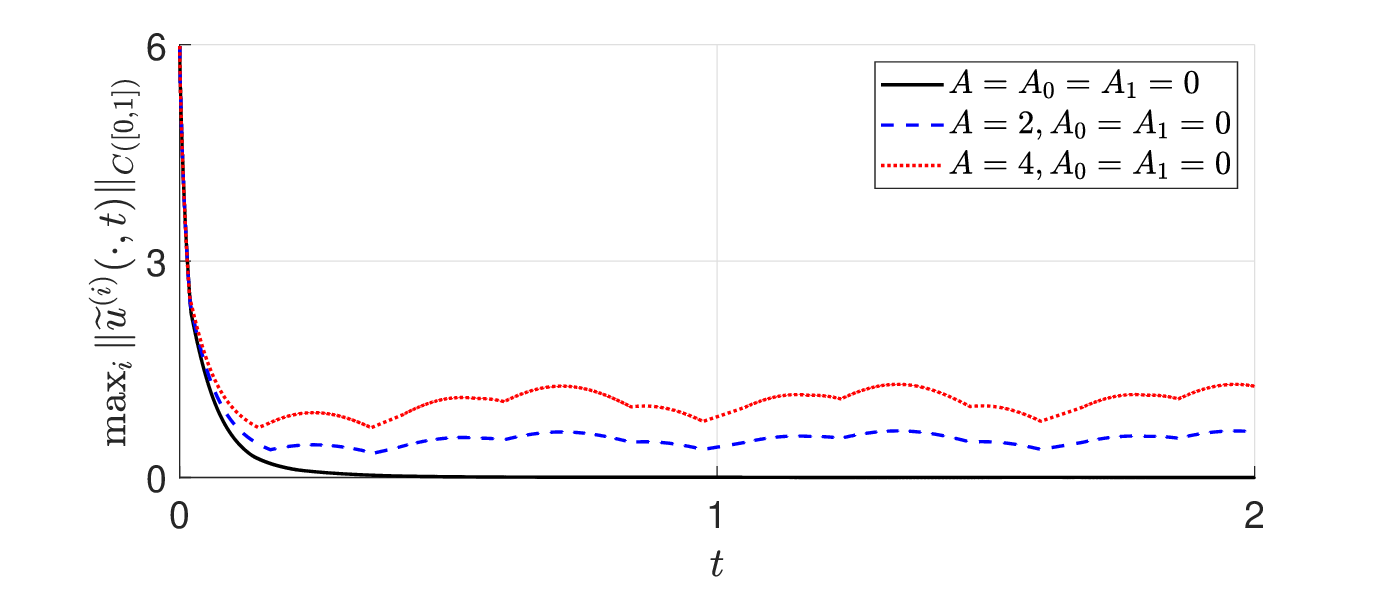}\\
													\includegraphics[scale= 0.3]{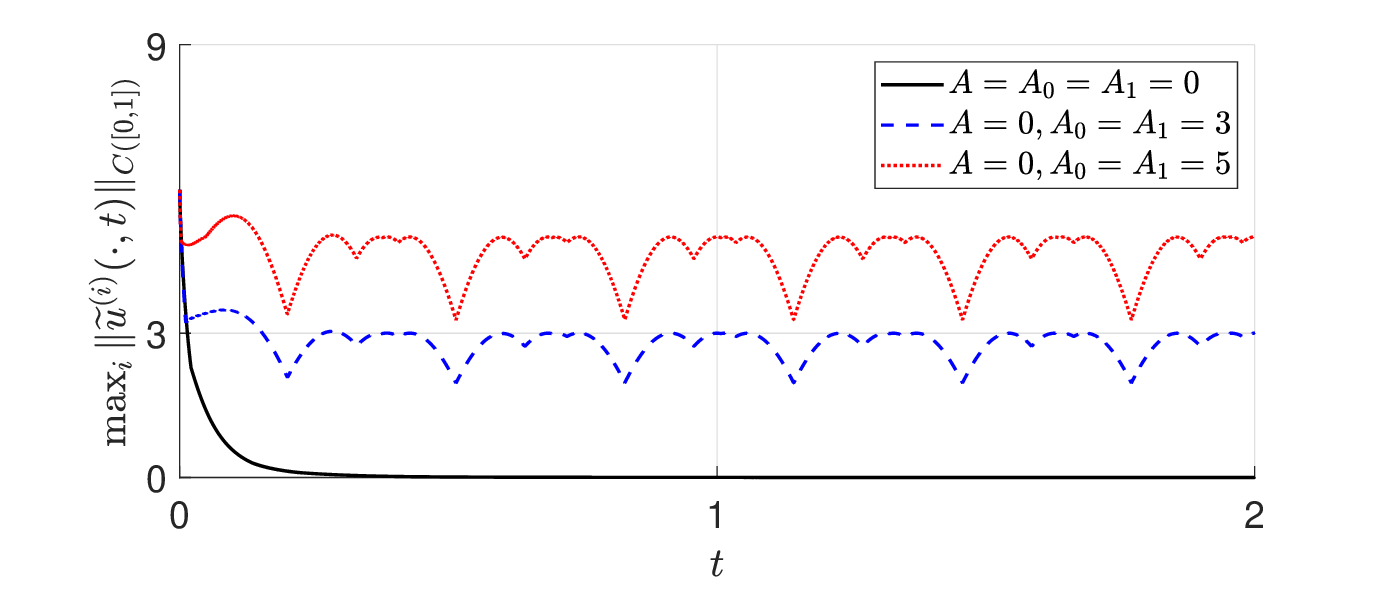}
													\caption{Evolution  of {$\max_{i}\|\widetilde u^{(i)}(\cdot,t)\|_{C([0,1])}$} for the tracking error  system~\eqref{widetildeu3} with different  {$f^{(i)}(x,t)$} (top) and {$d^{(i)}_0(t), d^{(i)}_1(t)$} (bottom) when $D_1=1$.}
													\label{fig6}
												\end{center}
											\end{figure}

\begin{figure}[htbp]
												\begin{center}
													\includegraphics[scale= 0.3]{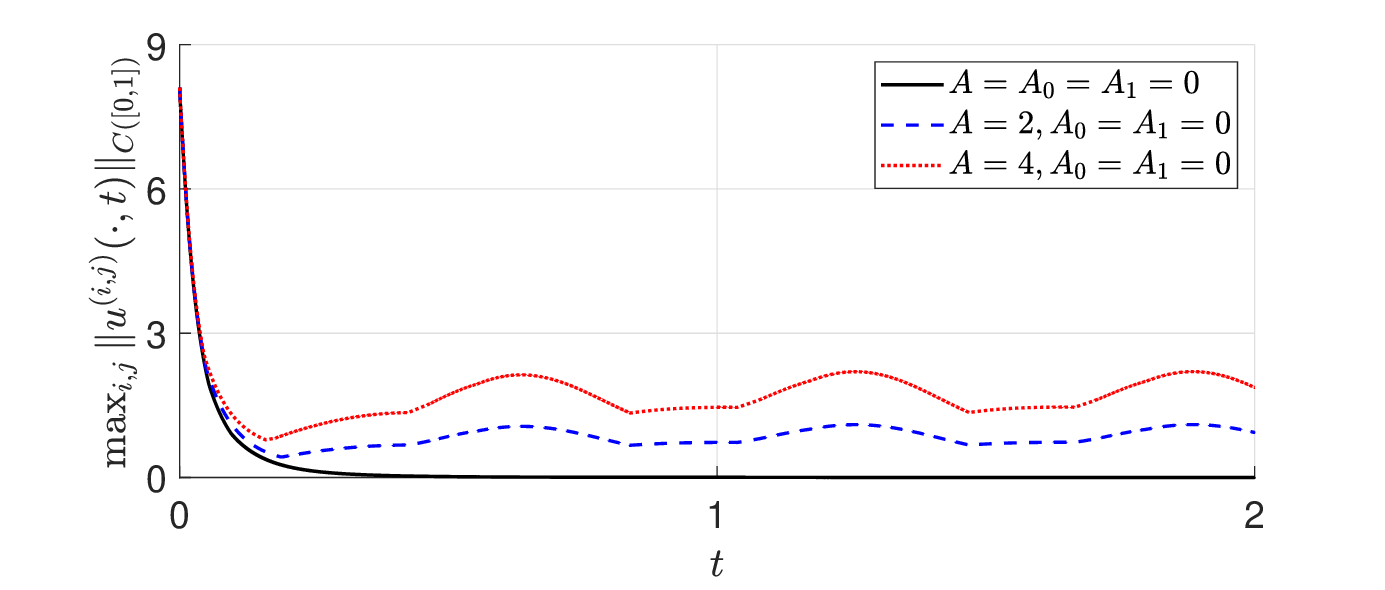}\\
													\includegraphics[scale= 0.3]{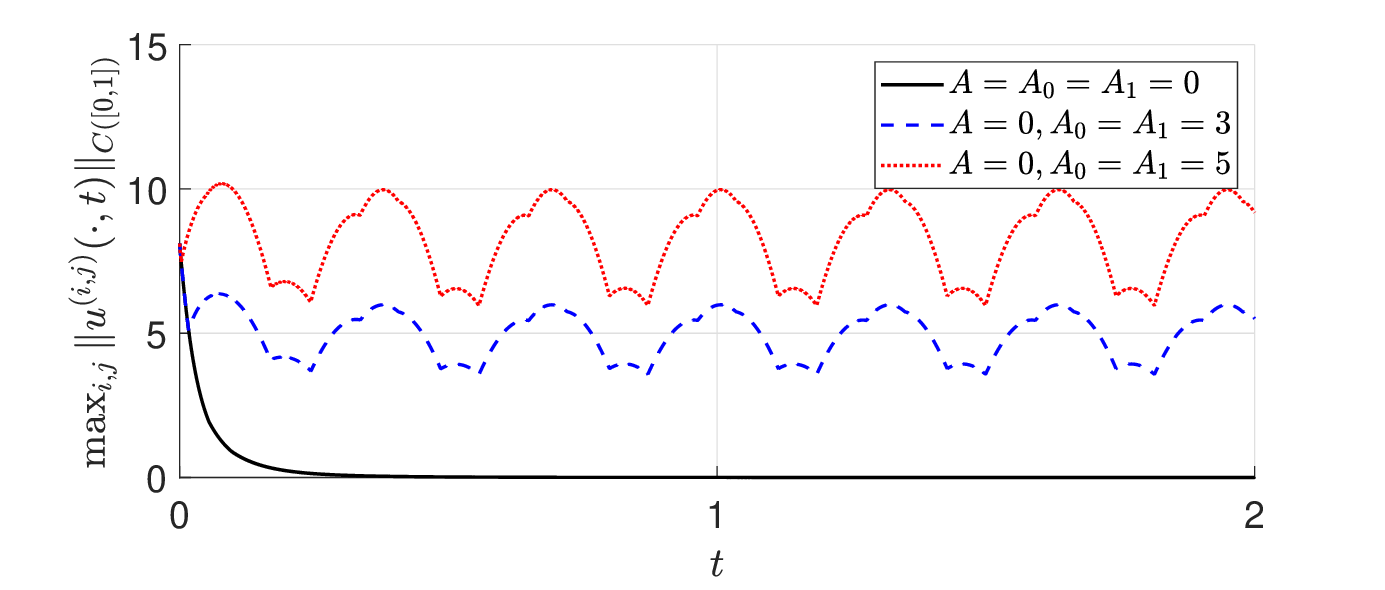}
													\caption{Evolution  of {$\max_{i,j}\|u^{(i,j)}(\cdot,t)\|_{C([0,1])}$} for the synchronization  error  system~\eqref{widetildeu-i-j} with different  {$f^{(i)}(x,t)$} (top) and {$d^{(i)}_0(t), d^{(i)}_1(t)$} (bottom) when $D_1=1$.}
													\label{fig7}
												\end{center}
											\end{figure}

											Under the control laws given by   \eqref{control lawv}, since all agents exhibit qualitatively similar behaviors, which can be inferred from the evolution of the tracking (or synchronization) errors shown in Fig.~\ref{fig6} (or Fig.~\ref{fig7}), we present in  Fig.~\ref{fig5} only  the evolution of the solution to  the first equation of the MAS~\eqref{muti-agent}       with $D_0=D_1=1$  under  different unobservable disturbances $f^{(i)}$, $d_0^{(i)}$, and $d_1^{(i)}$. From the first subfigure, it is observed that, in the absence of unobservable disturbances,   the state $u^{(1)} $ well tracks the reference trajectory $u^{\mathrm{ref}} $. The remaining subfigures indicate that, in the presence of these disturbances, the solutions remain  bounded, and their amplitudes decrease  as the disturbance magnitudes diminish.
											
											To further quantify the tracking performance, 	Fig.~\ref{fig6} shows the evolution of the maximum of all norms of   the tracking error   $\widetilde u^{(i)} $  in the presence of observable disturbances $q^{(i)}$  with $D_1=1$  and  different unobservable disturbances    $f^{(i)}$, $d_0^{(i)}$, and $d_1^{(i)}$. It can be seen that  $\max_{i}\|\widetilde u^{(i)}(\cdot,t)\|_{C([0,1])}$---and hence each  $ \|\widetilde u^{(i)}(\cdot,t)\|_{C([0,1])}$---decays to a small neighborhood of zero  when only   $q^{(i)}$ is present, while remaining bounded   when     $f^{(i)}$, $d_0^{(i)}$, and $d_1^{(i)}$ are also  involved. Moreover,  as the  amplitudes of   $f^{(i)}$, $d_0^{(i)}$, and  $d_1^{(i)}$ decrease, the  amplitudes  of   $\max_{i}\|\widetilde u^{(i)}(\cdot,t)\|_{C([0,1])}$ decreases accordingly. These results well illustrate the robustness of the tracking error  $\widetilde u^{(i)} $-system~\eqref{widetildeu3}.
											
											Figure~\ref{fig7} shows   the evolution of the maximum of all norms of  the synchronization error $u^{(i,j)}$  in the presence of observable disturbances $q^{(i)}$ with $D_1=1$  and   different unobservable disturbances  $f^{(i)}$, $d_0^{(i)}$, and $d_1^{(i)}$. It can be seen that $\max_{i,j}\allowbreak \|\widetilde u^{(i,j)}(\cdot,t)\|_{C([0,1])}$ decays to a small neighborhood of zero  in the absence of $f^{(i)}$, $d_0^{(i)}$, and $d_1^{(i)}$, while remaining bounded when $f^{(i)}$, $d_0^{(i)}$, and $d_1^{(i)}$ are involved. Moreover, as the amplitudes of $f^{(i)}$, $d_0^{(i)}$, and $d_1^{(i)}$ decrease, the amplitudes of $\max_{i,j}\allowbreak \|\widetilde u^{(i,j)}(\cdot,t)\|_{C([0,1])}$ also decrease, thereby validating the synchronization performance of the MAS and demonstrating its robustness.
											

											\begin{figure}[htbp]
												\begin{center}
													\includegraphics[scale= 0.3]{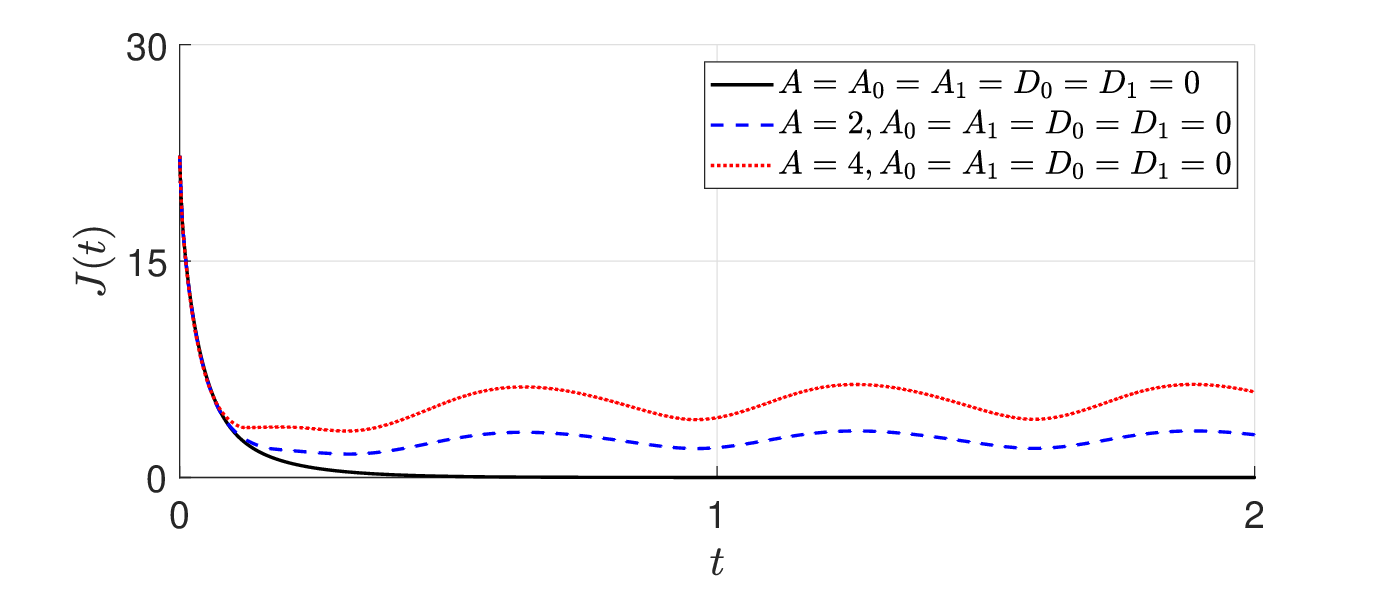}\\
													\includegraphics[scale= 0.3]{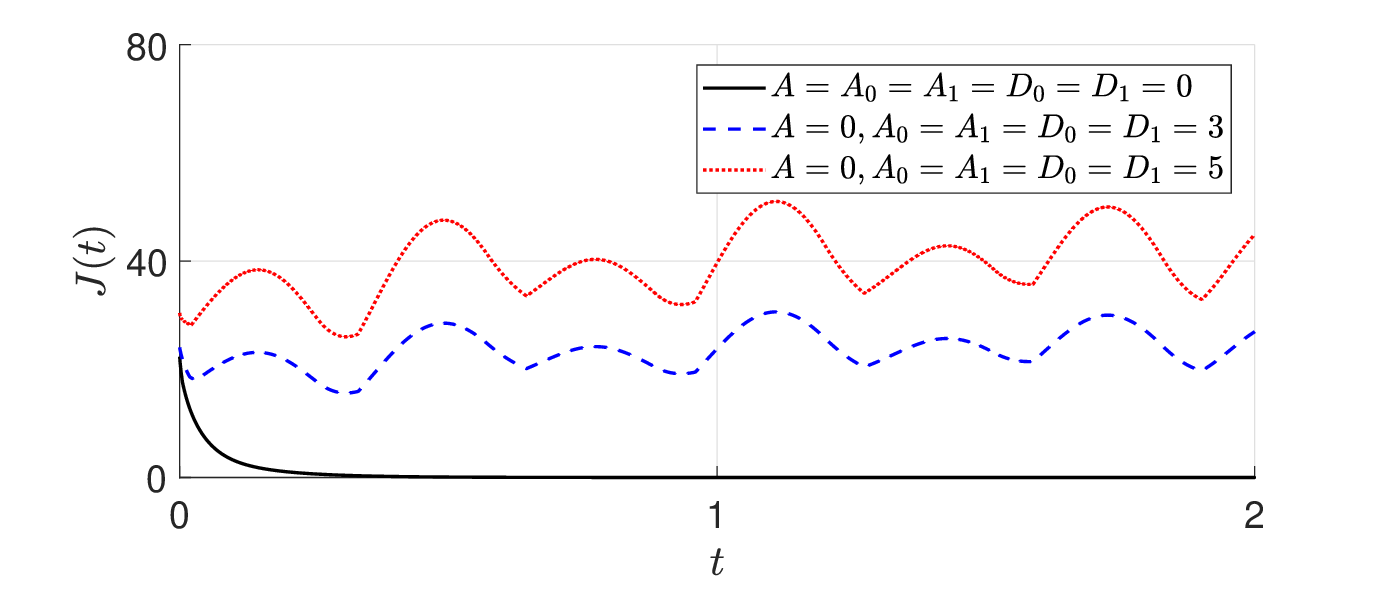}
													\caption{Evolution  of $J(t):=\max_i\Big(\left\| u^{(i)}(\cdot, t) \right\|_{C([0,1])} + \left\|  u^{\text{ref}}(\cdot, t) \right\|_{C([0,1])}  + \left\|  v^{(i)}(\cdot, t) \right\|_{C([0,1])} + \left\| \hat{q}^{(i)}(\cdot, t) \right\|_{C([0,1])}\Big)$ for the closed-loop  system composed of \eqref{muti-agent}--\eqref{control lawv} with different  {$f^{(i)}(x,t)$} (top) and {$d^{(i)}_0(t), d^{(i)}_1(t),r(t),q^{(i)}(t)$} (bottom).}
													\label{fig8}
												\end{center}
											\end{figure}

											Figure \ref{fig8} shows the evolution of $J(t):=\max_i\Big(\left\| u^{(i)}(\cdot, t) \right\|_{C([0,1])} + \left\|  u^{\text{ref}}(\cdot, t) \right\|_{C([0,1])}  + \left\|  v^{(i)}(\cdot, t) \right\|_{C([0,1])} + \left\| \hat{q}^{(i)}(\cdot, t) \right\|_{C([0,1])}\Big)$ for  the closed-loop system  composed of \eqref{muti-agent}--\eqref{control lawv} with different   $f^{(i)}$, $d_0^{(i)}$,   $d_1^{(i)}$, $r$, and $q^{(i)}$. It can be seen that $J(t)$ decays to a small neighborhood of zero in the absence of  $f^{(i)}$, $d_0^{(i)}$,   $d_1^{(i)}$, $r$, and $q^{(i)}$, while remaining   bounded when they are present. Furthermore, as the magnitudes of the disturbances decrease, the bound of $J(t)$ decreases accordingly. This verifies the ISS of  the closed-loop system  in the presence external signal input and disturbances.
											
											Overall, Fig. \ref{fig1}--\ref{fig8} demonstrate  that the proposed controllers   not only effectively reject  observable Dirichlet boundary disturbances, thereby  achieving reference tracking, but also ensure  robust synchronization among agents and the ISS of the closed-loop system in the presence of various  unobservable disturbances.

											%
											%
											
											\section{Conclusion}\label{Sec. VII}
											For the synchronization of  MASs  governed by parabolic PDEs with multiple types of disturbances simultaneously,  robust  synchronization controllers were designed by using the  reference
										signal and local output information of the agents,  enabling  all agents to track a prescribed reference trajectory while rejecting observable Dirichlet boundary disturbances  and ensuring robustness against the remaining unobservable disturbances without any prior information on their  magnitude, dynamics, and structure.
										While most existing studies    have focused on a single disturbance type or  are limited to cases where disturbances are either fully observable or partially known,	the coexistence of observable Dirichlet boundary disturbances and unobservable disturbances---including in-domain disturbances and Dirichlet-Robin boundary disturbances without prior information---poses   significant challenges for stability assessment. This work addressed these challenges by using the GLM, {which allows}
										systematically handling both observable and unobservable
										disturbances.  Using this approach, we established robustness properties for the estimation, tracking, and synchronization error systems, along with the ISS of the closed-loop system, thus quantifying the influence of disturbances on synchronization performance.
										The control and stability analysis framework developed in this paper will be extended in future work to address more challenging multi-agent control problems, including  multi-agent formation control and systems described by  complex PDEs under general forms, all in
										the presence of {various  types of disturbances}. 


\begin{thebibliography}{10}
	
	\bibitem{Aguilar2021}
	L.~Aguilar, Y.~Orlov, and A.~Pisano.
	\newblock Leader-follower synchronization and {ISS} analysis for a network of
	boundary-controlled wave {PDE}s.
	\newblock {\em IEEE Control Systems Letters}, 5(2):683--688, 2021.
	
	\bibitem{chen2020SCL}
	Y.~Chen, Z.~Zuo, and Y.~Wang.
	\newblock Bipartite consensus for a network of wave equations with time-varying
	disturbances.
	\newblock {\em Systems \& Control Letters}, 136, 2020.
	\newblock 104604.
	
	\bibitem{chen2021auto}
	Y.~Chen, Z.~Zuo, and Y.~Wang.
	\newblock Bipartite consensus for a network of wave {PDE}s over a signed
	directed graph.
	\newblock {\em Automatica}, 129, 2021.
	\newblock 109640.
	
	\bibitem{chen2021JFI}
	Z.~Chen, J.~Yang, and X.~Zong.
	\newblock Leader-follower synchronization controller design for a network of
	boundary-controlled wave {PDE}s with structured time-varying perturbations
	and general disturbances.
	\newblock {\em Journal of the Franklin Institute}, 358(1):834--855, 2021.
	
	\bibitem{Demetriou2013scl}
	M.~A. Demetriou.
	\newblock Synchronization and consensus controllers for a class of parabolic
	distributed parameter systems.
	\newblock {\em Systems \& Control Letters}, 62(1):70--76, 2013.
	
	\bibitem{demetriou2018}
	M.~A. Demetriou.
	\newblock Design of adaptive output feedback synchronizing controllers for
	networked {PDE}s with boundary and in-domain structured perturbations and
	disturbances.
	\newblock {\em Automatica}, 90:220--229, 2018.
	
	\bibitem{deutscher2022}
	J.~Deutscher.
	\newblock Robust cooperative output regulation for a network of parabolic {PDE}
	systems.
	\newblock {\em IEEE Transactions on Automatic Control}, 67(1):451--459, 2022.
	
	\bibitem{Evans2010}
	L.~C. Evans.
	\newblock {\em Partial {D}ifferential {E}quations}.
	\newblock American Mathematical Society, Providence, Rhode Island, 2010.
	
	\bibitem{Frihauf2011}
	P.~Frihauf and M.~Krstic.
	\newblock Leader-enabled deployment onto planar curves: A {PDE}-based approach.
	\newblock {\em IEEE Transactions on Automatic Control}, 56(8):1791--1806, 2011.
	
	\bibitem{Ghods2012}
	N.~Ghods and M.~Krstic.
	\newblock Multiagent deployment over a source.
	\newblock {\em IEEE Transactions on Control Systems Technology},
	20(1):277--285, 2012.
	
	\bibitem{guo2015TAC}
	B.~Z. Guo and H.~C. Zhou.
	\newblock The active disturbance rejection control to stabilization for
	multi-dimensional wave equation with boundary control matched disturbance.
	\newblock {\em IEEE Transactions on Automatic Control}, 60(1):143--157, 2015.
	
	\bibitem{Han2009TIE}
	J.~Han.
	\newblock From {PID} to active disturbance rejection control.
	\newblock {\em IEEE Transactions on Industrial Electronics}, 56(3):900--906,
	2009.
	
	\bibitem{he2018IET}
	P.~He.
	\newblock Consensus of uncertain parabolic {PDE} agents via adaptive
	unit-vector control scheme.
	\newblock {\em IET Control Theory \& Applications}, 12(18):2488--2494, 2018.
	
	\bibitem{Konduri2013}
	S.~Konduri, P.~R. Pagilla, and S.~Darbha.
	\newblock Vehicle formations using directed information flow graphs.
	\newblock In {\em 2013 American Control Conference (ACC)}, pages 3045--3050,
	Washington, DC, USA, 2013.
	
	\bibitem{Kumar2020}
	S.~R. Kumar and D.~Mukherjee.
	\newblock Cooperative salvo guidance using finite-time consensus over directed
	cycles.
	\newblock {\em IEEE Transactions on Aerospace and Electronic Systems},
	56(2):1504--1514, 2020.
	
	\bibitem{li2023ND}
	L.~Li and J.~Liu.
	\newblock Consensus tracking control and vibration suppression for nonlinear
	mobile flexible manipulator multi-agent systems based on {PDE} model.
	\newblock {\em Nonlinear Dynamics}, 111(4):3345--3359, 2023.
	
	\bibitem{liu2025NNLS}
	Y.~J. Liu, X.~Shang, L.~Tang, and S.~Zhang.
	\newblock Finite-time consensus adaptive neural network control for nonlinear
	multiagent systems under {PDE} models.
	\newblock {\em IEEE Transactions on Neural Networks and Learning Systems},
	36(4):6218--6228, 2025.
	
	\bibitem{Marshall2004}
	J.~Marshall, M.~Broucke, and B.~Francis.
	\newblock Formations of vehicles in cyclic pursuit.
	\newblock {\em IEEE Transactions on Automatic Control}, 49(11):1963--1974,
	2004.
	
	\bibitem{mcarthur2007}
	S.~D.~J. McArthur, E.~M. Davidson, V.~M. Catterson, A.~L. Dimeas, N.~D.
	Hatziargyriou, F.~Ponci, et~al.
	\newblock Multi-agent systems for power engineering applications---part {I}:
	Concepts, approaches and technical challenges.
	\newblock {\em IEEE Transactions on Power Systems}, 22(4):1743--1752, 2007.
	
	\bibitem{Meurer2011}
	T.~Meurer and M.~Krstic.
	\newblock Finite-time multi-agent deployment: A nonlinear {PDE} motion planning
	approach.
	\newblock {\em Automatica}, 47(11):2534--2542, 2011.
	
	\bibitem{Mironchenko:2023b}
	A.~Mironchenko.
	\newblock Input-to-state stability of distributed parameter systems.
	\newblock arXiv:2302.00535.
	
	\bibitem{Mironchenko2020}
	A.~Mironchenko and C.~Prieur.
	\newblock Input-to-state stability of infinite-dimensional systems: Recent
	results and open questions.
	\newblock {\em SIAM Review}, 62(3):529--614, 2020.
	
	\bibitem{Morstyn2016}
	T.~Morstyn, B.~Hredzak, and V.~G. Agelidis.
	\newblock Cooperative multi-agent control of heterogeneous storage devices
	distributed in a {DC} microgrid.
	\newblock {\em IEEE Transactions on Power Systems}, 31(4):2974--2986, 2016.
	
	\bibitem{Orlov2016IFAC}
	Y.~Orlov, A.~Pilloni, A.~Pisano, and E.~Usai.
	\newblock Consensus-based leader-follower tracking for a network of perturbed
	diffusion {PDE}s via local boundary interaction.
	\newblock {\em IFAC-PapersOnLine}, 49(8):228--233, 2016.
	
	\bibitem{pazy1983}
	A.~Pazy.
	\newblock {\em Semigroups of Linear Operators and Applications to Partial
		Differential Equations}.
	\newblock Springer, New York, NY, 1983.
	
	\bibitem{Pilloni2016}
	A.~Pilloni, A.~Pisano, Y.~Orlov, and E.~Usai.
	\newblock Consensus-based control for a network of diffusion {PDE}s with
	boundary local interaction.
	\newblock {\em IEEE Transactions on Automatic Control}, 61(9):2708--2713, 2016.
	
	\bibitem{Qi2019ijc}
	J.~Qi, S.-X. Tang, and C.~Wang.
	\newblock Parabolic {PDE}-based multi-agent formation control on a cylindrical
	surface.
	\newblock {\em International Journal of Control}, 92(1):77--99, 2019.
	
	\bibitem{Qi2015}
	J.~Qi, R.~Vazquez, and M.~Krstic.
	\newblock Multi-agent deployment in 3-{D} via {PDE} control.
	\newblock {\em IEEE Transactions on Automatic Control}, 60(4):891--906, 2015.
	
	\bibitem{Qi2019}
	J.~Qi, S.~Wang, J.-A. Fang, and M.~Diagne.
	\newblock Control of multi-agent systems with input delay via {PDE}-based
	method.
	\newblock {\em Automatica}, 106:91--100, 2019.
	
	\bibitem{qi2018}
	J.~Qi, J.~Zhang, and Y.~Ding.
	\newblock Wave equation-based time-varying formation control of multiagent
	systems.
	\newblock {\em IEEE Transactions on Control Systems Technology},
	26(5):1578--1591, 2018.
	
	\bibitem{Qiu2023TSMCS}
	Q.~Qiu and H.~Su.
	\newblock Exponential consensus of multiagent systems based on nonlinear
	parabolic {PDE}s via intermittent boundary control.
	\newblock {\em IEEE Transactions on Systems, Man, and Cybernetics: Systems},
	53(8):4833--4842, 2023.
	
	\bibitem{Rogge2008}
	J.~A. Rogge and D.~Aeyels.
	\newblock Vehicle platoons through ring coupling.
	\newblock {\em IEEE Transactions on Automatic Control}, 53(6):1370--1377, 2008.
	
	\bibitem{Tang2017ijc}
	S.~X. Tang, J.~Qi, and J.~Zhang.
	\newblock Formation tracking control for multi-agent systems: A wave-equation
	based approach.
	\newblock {\em International Journal of Control, Automation and Systems},
	15:2704--2713, 2017.
	
	\bibitem{wang2024auto}
	S.~Wang, M.~Diagne, and J.~Qi.
	\newblock Delay-adaptive compensation for 3-{D} formation control of
	leader-actuated multi-agent systems.
	\newblock {\em Automatica}, 164, 2024.
	\newblock 111645.
	
	\bibitem{Yan2024TNSE}
	X.~Yan, K.~Li, C.~Yang, J.~Zhuang, and J.~Cao.
	\newblock Consensus of fractional-order multi-agent systems via observer-based
	boundary control.
	\newblock {\em IEEE Transactions on Network Science and Engineering},
	11(4):3370--3382, 2024.
	
	\bibitem{Yang2017}
	C.~Yang, H.~He, T.~Huang, A.~Zhang, J.~Qiu, J.~Cao, and X.~Li.
	\newblock Consensus for non-linear multi-agent systems modelled by {PDE}s based
	on spatial boundary communication.
	\newblock {\em IET Control Theory \& Applications}, 11(17):3196--3200, 2017.
	
	\bibitem{yao2024}
	Q.~Yao, Q.~Li, C.~Huang, and H.~Jahanshahi.
	\newblock {PDE}-based distributed adaptive fault-tolerant attitude consensus of
	multiple flexible spacecraft.
	\newblock {\em Aerospace Science and Technology}, 146, 2024.
	\newblock 108934.
	
	\bibitem{Yu2019}
	Y.~Yu, Z.~Li, X.~Wang, and L.~Shen.
	\newblock Bearing-only circumnavigation control of the multi-agent system
	around a moving target.
	\newblock {\em IET Control Theory \& Applications}, 13(17):2747--2757, 2019.
	
	\bibitem{Zhan2025TAC}
	J.~Zhan, L.~Zhang, and J.~Qiao.
	\newblock Boundary consensus of networked hyperbolic systems of conservation
	laws.
	\newblock {\em IEEE Transactions on Automatic Control}, 70(8):4989--5004, 2025.
	
	\bibitem{ZhangIJRNC}
	H.~Zhang, J.~Li, T.~Wang, Q.~Zeng, and H.~Yan.
	\newblock {PDE}-based event-triggered containment control of multi-agent
	systems with input delay under spatial boundary communication.
	\newblock {\em International Journal of Robust and Nonlinear Control},
	33(2):753--766, 2023.
	
	\bibitem{zhang2024auto}
	J.~Zhang, R.~Vazquez, J.~Qi, and M.~Krstic.
	\newblock Multi-agent deployment in 3-{D} via reaction¨Cdiffusion system with
	radially-varying reaction.
	\newblock {\em Automatica}, 161, 2024.
	\newblock 111491.
	
	\bibitem{Zhang2024TPDS}
	S.~Zhang, L.~Tang, and Y.~J. Liu.
	\newblock Adaptive neural control for a network of parabolic {PDE}s with
	event-triggered mechanism.
	\newblock {\em IEEE Transactions on Parallel and Distributed Systems},
	35(7):1320--1330, 2024.
	
	\bibitem{Zhao2024TNNLS}
	W.~Zhao, Y.~Liu, and X.~Yao.
	\newblock {PDE}-based boundary adaptive consensus control of multiagent systems
	with input constraints.
	\newblock {\em IEEE Transactions on Neural Networks and Learning Systems},
	35(9):12617--12626, 2024.
	
	\bibitem{Zheng2018}
	J.~Zheng and G.~Zhu.
	\newblock Input-to-state stability with respect to boundary disturbances for a
	class of semi-linear parabolic equations.
	\newblock {\em Automatica}, 97(8):271--277, 2018.
	
	\bibitem{Zheng2024}
	J.~Zheng and G.~Zhu.
	\newblock Global {{ISS}} for the viscous {{B}}urgers' equation with
	{{D}}irichlet boundary disturbances.
	\newblock {\em IEEE Transactions on Automatic Control}, 69(10):7174--7181,
	2024.
	
	\bibitem{Zheng2025}
	J.~Zheng and G.~Zhu.
	\newblock Generalized {Lyapunov} functionals for the input-to-state stability
	of infinite-dimensional systems.
	\newblock {\em Automatica}, 172, 2025.
	\newblock 112005.
	
	\bibitem{zong2021}
	X.~Zong, J.~Zheng, J.~Wei, and J.~Yang.
	\newblock Synchronizing controller design for networked heat equations with
	boundary structured time-varying perturbations and general disturbances.
	\newblock {\em Journal of Process Control}, 107:94--102, 2021.
	
\end{thebibliography}
\end{document}